\documentclass[12pt]{article}
\usepackage{amsfonts}
\usepackage{bbm}
\usepackage{bm}
\usepackage{amsmath}
\usepackage{amscd}
\usepackage{eufrak}

\newcommand{\bmat}{\left(\begin{array}}
\newcommand{\emat}{\end{array}\right)}

\def\gtrsim{\mathrel{\raise.3ex\hbox{$>$\kern-.75em\lower1ex\hbox{$\sim$}}}}

\def\om{\omega}

\def\-{\hphantom{-}}
\def\ov{\overline}
\def\s2{\frac{1}{\sqrt2}}

\def\oh{\frac{1}{2}}
\def\beq{\begin{equation}}
\def\eeq{\end{equation}}
\def\beqa{\begin{eqnarray}}
\def\eeqa{\end{eqnarray}}

\def\T{{\rm T}}
\def\Z{{\mathbb Z}}

\def\ba{{\bar a}}
\def\bb{{\bar b}}
\def\bh{{\bar h}}
\def\bg{{\bar g}}

\def\mg{m_{3/2}}
\def\mg2{m^2_{3/2}}

\def\Dsl{\,\raise.15ex\hbox{/}\mkern-13.5mu D} %this one can be subscripted

\newcommand{\mathsmaller}[1]{\mbox{\footnotesize$#1$}}

\def\bF{{\bm{F}}}
\def\bR{{\bm{R}}}
\def\aneq{\not=}

\begin{document}
\pagestyle{plain}

%----------------------------------------------------------------------%
%  numbering equations with section number
%----------------------------------------------------------------------%
\makeatletter
\@addtoreset{equation}{section}
\makeatother
\renewcommand{\theequation}{\thesection.\arabic{equation}}
%----------------------------------------------------------------------%
%  title page
%----------------------------------------------------------------------%
\pagestyle{empty}
\rightline{ CPHT-RR083.1108}
\begin{center}
\LARGE{Flux algebra, Bianchi identities and Freed-Witten anomalies in F-theory compactifications\\[10mm]}
\large{ G. Aldazabal${}^{a,c}$, P. G. C\'amara${}^b$, J.  A.
Rosabal${}^c$
 \\[6mm]}
\small{
${}^a$Centro At\'omico Bariloche, ${}^c$Instituto Balseiro
(CNEA-UNC) and CONICET. \\[-0.3em]
8400 S.C. de Bariloche, Argentina. \\
${}^b$Centre de Physique Th\'eorique\footnote{Unit\'e mixte
du CNRS, UMR 7644.}, Ecole Polytechnique,\\[-0.3em]
F-91128 Palaiseau, France.
\\[1cm]}
\small{\bf Abstract} \\[0.5cm]
\end{center}

{\small

We discuss the structure of 4D gauged supergravity algebras
corresponding to globally non-geometric compactifications of
F-theory, admitting a local geometric description in terms of 10D
supergravity. By starting with the well known algebra of gauge
generators associated to non-geometric type IIB fluxes, we derive
a full algebra containing all, closed RR and NSNS, geometric and
non-geometric dual fluxes. We achieve this generalization by a
systematic application of $SL(2,\Z)$ duality transformations and
by taking care of the spinorial structure of the fluxes. The
resulting algebra encodes much information about the higher
dimensional theory. In particular, tadpole equations and Bianchi
identities are obtainable as Jacobi identities of the algebra.
When a sector of magnetized $(p,q)$ 7-branes is included, certain
closed axions are gauged by the $U(1)$ transformations on the
branes. We indicate how the diagonal gauge generators of the
branes can be incorporated into the full algebra, and show that
Freed-Witten constraints and tadpole cancellation conditions for
$(p,q)$ 7-branes can be described as Jacobi identities satisfied
by the algebra mixing bulk and brane gauge generators. }

\newpage
%----------------------------------------------------------------------%
%  Resetting of counters
%----------------------------------------------------------------------%
\setcounter{page}{1}
\pagestyle{plain}
\renewcommand{\thefootnote}{\arabic{footnote}}
\setcounter{footnote}{0}
%----------------------------------------------------------------------%
%  Paper begins
%----------------------------------------------------------------------%

\tableofcontents

\section{Introduction}
\label{seci} The presence of a variety of $p$-forms in ten
dimensional superstring theories opens up the possibility of
considering compactifications to lower dimensions where some
background fluxes are turned on for their corresponding field
strengths. In recent years these flux compactifications
\cite{Grana:2005jc} (which include  the so-called
\emph{geometric fluxes}, deforming the internal compactification
manifold), have been studied from different perspectives. A
particularly appealing consequence is that fields -- which would
be moduli in the absence of fluxes -- acquire a potential
\cite{gvw} and, thus, vacuum degeneracy can be lifted. This is
particularly important for understanding the breaking of
supersymmetry.

The effective four dimensional field theories resulting from
superstring flux compactifications are described by gauged
supergravities, namely, deformations of ordinary abelian
supergravity theories where the deformation parameters (gaugings)
correspond to the quantized fluxes. Matter fields become charged
with respect to gauge fields which, generically, are  non-abelian.
A nice example of this situation is provided by Scherk-Schwarz
compactifications \cite{Scherk:1979zr} on twisted tori, where the isometries of the
compact manifold become structure constants of the four
dimensional gauge fields.

However, the connection between a string compactification and the gauged
supergravity effective theory is quite subtle. For
instance, it is known that orientifold compactifications of type
IIA and type IIB 10D supergravity actions, in the presence of
the corresponding background fluxes,  lead to different superpotentials.
Since both superstrings theories are connected through mirror
symmetry, it was suggested in \cite{stw,acfi} that new fluxes
should be included in order for the full superpotentials to match.
Similarly, more fluxes can be suggested by invoking type IIB
S-duality, M-theory or heterotic/type I S-duality \cite{acfi}.
Hence, the resulting four dimensional gauged supergravity theory
incorporates information about the stringy aspects of the starting
configuration. Generically, it is not the reduction of a ten
dimensional effective supergravity action.

Fluxes are strongly constrained by consistency conditions like
Bianchi identities or tadpole cancellation equations. Part of such
constraints can be derived from the ten dimensional effective
supergravity. However, other necessary constraints, related to
quantities encoding stringy aspects of the starting configuration,
must be inferred by using string duality arguments. This is a
particularly relevant issue when  the study of moduli
stabilization is addressed. Certain candidates for scalar potential minima
could be actually forbidden by these constraints.

In this paper we take the orientifold framework of Ref.
\cite{acfi}. There, type IIB orientifold compactifications plus a
$\mathbb{Z}_2\times\mathbb{Z}_2$ projection ensuring tori
factorization is considered. In particular it is shown that  $2^7$
dual fluxes are needed to ensure duality invariance and, moreover,
these are arranged into spinor representations of $SL(2,\Z)^7$. We
focus on $2^6$ of these fluxes and argue that they describe
globally non-geometric compactifications of F-theory, which admit
a geometric local description in terms of 10D supergravity. Hence,
in Section \ref{sec1}, we generalize the construction of
\cite{stw} to allow for type IIB $SL(2,\mathbb{Z})$
transformations, apart from T-dualities, in the transition
functions which glue different local patches. The canonical
example is given by F-theory compactification on a K3 \cite{sen},
where some of the 7-branes have been traded by closed string
fluxes, leading to an axion-dilaton which is not neutral under
global monodromies. The resulting compact space is generically an
U-fold \cite{ufolds1,ufolds} (see also \cite{reid}).

Recently, the application of F-theory to model building has been
explored \cite{vafaf,masf,masf1,masf2,masf3,masf4}, showing
interesting features, such as top Yukawa couplings of the order of
the gauge coupling constant or the presence of exceptional gauge
groups, suitable for GUT model building. On the other side,
globally non-geometric compactifications may lead to moduli
stabilization\footnote{Although this stabilization manifests
perturbatively in the four dimensional effective theory, we prefer
not to use the term \emph{perturbative}, since in our case the ten
dimensional uplift is generically non-perturbative.}
\cite{stw,acfi}, or induce supersymmetric mass terms ($\mu$-terms)
in the worldvolume of D3 and D7-branes
\cite{soft1,soft2,soft3,soft4,soft5,soft6}. Remarkably,
supersymmetric Minkowski vacua with all closed string moduli
stabilized were found in \cite{vafanongeo,eran} at the level of
the 4D effective supergravity.

The aim of this paper is to provide the necessary tools for
combining both setups, F-theory configurations with exotic
7-branes and globally non-geometric compactifications, in a
consistent manner. For that, we exploit the connection of string
compactifications to gaugings in the four dimensional theory, and
propose a systematic way to study the constraints that background
fluxes must obey.

In a general compactification of F-theory there are three possible
origins of four dimensional vector fields which can be gauged:
\begin{enumerate}
\item Kaluza-Klein vectors, arising from dimensional reduction of
the metric and the NSNS/RR forms.

\item 7-brane gauge symmetries, arising at codimension two loci
where the fiber, parametrizing the axion-dilaton dependence,
degenerates into an ADE singularity. At points of the moduli space
admitting an orbifold description, some of these vectors
correspond to twisted states.

\item 3-brane gauge symmetries, completely localized along the
compact directions.
\end{enumerate}
In this work we analyze the gauged supergravity algebras
associated to the first two classes of vector fields, and comment
also on the constraints arising from 3-branes, when the latter
sector is trivial. Our starting point is the algebra satisfied by
a small set of gauge generators whose ten dimensional origin is
well understood. A further systematic application of $SL(2,\Z)^7$
transformations then leads to a complete algebra for Kaluza-Klein
vectors, where the full set of fluxes is generated as structure
constants. Jacobi identities satisfied by gauge generators lead to
constraints on the allowed fluxes. Interestingly enough, from this
four dimensional point of view, RR tadpole cancellation
requirements and  Bianchi identities appear at the same footing.
We derive the tadpole cancellation requirements for $(p,q)$
7-branes and show that some Jacobi identities impose stronger
restrictions than other previously considered in the literature.
This study is performed in Section \ref{sec2}.

In Section \ref{sec4} we address the inclusion of the 7-brane
sector. In principle, chiral configurations of branes can  be
described in the gauging formalism. One way is to start from e.g.
type I with D9 branes, and then magnetize them. The magnetization
simply amounts to introducing some worldvolume fluxes, which can
be described as gaugings. Due to Chern-Simons couplings on the
volume of the branes, sourced by the presence of magnetic fields,
certain RR axionic scalars shift under D-brane $U(1)$ gauge
symmetries.

From different points of view, it has been put forward that some
D-brane configurations are not allowed for when bulk fluxes are
switched on. The allowed brane configurations must satisfy some
non-trivial generalized Freed-Witten \cite{Freed:1999vc}
constraints, indicating a sort of mutual consistency between open
and closed string fluxes. For instance, in \cite{cfi} it was
proposed to understand Freed-Witten constraints as the
requirements needed to ensure invariance of the effective
superpotential under shifts of the $U(1)$ gauged axionic scalars.
Here we propose to study this interrelation between open and
closed string fluxes, from the point of view of the complete
algebra describing both, bulk and worldvolume gaugings.  The idea
is that Freed-Witten constraints could arise as Jacobi identities
for such algebra. We find that this is indeed the case, at least
for the subset of fluxes that we are able to manage. We derive in
this way the Freed-Witten conditions that  $(p,q)$ 7-branes have
to satisfy in non-geometric compactifications of F-theory.

In Section \ref{sec3} we discuss the connection between
$\mathcal{N}=4$ supergravity actions and $\mathcal{N}=1$
superstring compactifications. We expect our string
compactification to lead in four dimensions to a truncation of an
$\mathcal{N}=4$ gauged supergravity theory, due to the presence of
fluxes and the orientifold projection. Indeed, we show that a
precise dictionary between the $\mathcal{N}=4$ orbifold truncated
gauged algebra and the one derived in Section \ref{sec2} can be
established. Jacobi identities do match quadratic constraints on
the possible gaugings, required for consistency of gauged
supergravity theories.

Finally, Section \ref{sec5} contains some conclusions and an
outlook. We collect in the Appendix different tables containing
fluxes and their transformation properties.

%%%%%%%%%%%%%%%%%%%%%%%%%%%%%%%%%%%%%%%%%%%%%%%%%%%%%%%%%%%%%%%%%%%%%%%

\section{Basics}
\label{sec1}

\subsection{Type IIB compactifications with non-geometric fluxes}
\label{sec21}

We consider type IIB orientifold compactifications on
$T^6/[\Omega_P(-1)^{F_L}\sigma]$, where $\Omega_P$ is the
worldsheet parity operator, $(-1)^{F_L}$ is the space-time
fermionic number for the left-movers and $\sigma$ is an involution
acting on the K\"ahler form and the holomorphic 3-form,
respectively as $\sigma(J)=J$ and $\sigma(\Omega)=-\Omega$. In
general there can be O3-planes spanning the space-time directions.

For simplicity we also assume  an underlying
$\mathbb{Z}_2\times\mathbb{Z}_2$ symmetry, so the 6-torus is
factorized. Prior to the inclusion of background fluxes and
stringy monodromies, the metric of the internal torus reads
\begin{equation}
ds^2=\sum_{k=1}^3\frac{\textrm{Re }T_k}{\textrm{Re
}U_k}|dx^k+iU_kdx^{k+3}|^2\ ,\label{metrica}
\end{equation}
where $U_k$ are the complex structure moduli and $T_k$ are the
K\"ahler moduli, defined in terms of the RR 4-form and the
K\"ahler 2-form
\begin{equation}
\mathcal{J}_c\equiv C_4+\frac{i}{2}e^{-\phi}J\wedge J\ =
i\sum_{i=1}^3T_i\tilde\omega_i\ .\label{c4}
\end{equation}
Here, $\tilde\omega_i$ denotes a basis of invariant 4-forms.

The axion-dilaton modulus is given by
\begin{equation}
S=e^{-\phi}+iC_0\ ,\label{s}
\end{equation}
with $C_0$ the RR 0-form. In the presence of 7-branes, it is a
holomorphic function of the internal coordinates. The four
dimensional field, $S$, is then defined as the constant mode of
the ten dimensional fields.

The seven moduli of the compactification, $(S,T_k,U_k)$,
$k=1,2,3$, span a K\"ahler manifold with potential
\begin{equation}
K=-\textrm{log }(S+\bar S)-\sum_{k=1}^3\textrm{log}[(T_k+\bar
T_k)(U_k+\bar U_k)]\ .\label{kahler}
\end{equation}

The orientifold involution allows for two types of BPS
objects\footnote{Strictly speaking, also D9-branes with suitably
magnetization are allowed in certain regions of the moduli space.
Here, however, we only focus on 3 and 7-branes.\label{d9}}:
D3-branes and $(p,q)$ $7_k$-branes, with RR charge $p$ and NSNS
charge $q$. The first ones span the four space-time dimensions,
whereas the latter wrap also a (2,2)-cycle, $[\tilde \omega_k]$,
in the internal 6-torus.

Consistently with the orientifold involution, it is possible to
consider in addition non-trivial backgrounds for the RR and NSNS
3-forms, $F_3$ and $H_3$, satisfying a BPS-like condition
\cite{gkp}. The fluxes induce a deformation on the moduli space
which, from the four dimensional perspective, can be effectively
parameterized in terms of the superpotential \cite{gvw}
\begin{equation}
W=\int (F_3-iSH_3)\wedge \Omega\ .\label{gvw}
\end{equation}
Due to the Chern-Simons coupling
\begin{equation}
\int C_4\wedge H_3\wedge F_3\ ,
\end{equation}
present in the type IIB supergravity action, the fluxes induce a
smeared charge of D3-brane which, for consistency, must be
cancelled by the effect of localized sources. This leads to the
condition,
\begin{equation}
\frac{1}{2\cdot3!}\tilde{F}^{mnp}H_{mnp}=N_{D3/O3}\
,\label{d3tadpole}
\end{equation}
where $N_{D3/O3}$ denotes the total D3-brane charge associated
with localized objects and,
\begin{equation}
\tilde{F}^{ijk}\equiv \frac{1}{3!}\epsilon^{ijkopq}F_{opq}\
.\label{tildeF}
\end{equation}
Since D3-branes are related to 3-form fluxes through brane/flux
transmutation \cite{ks,gtrans}, the D3 tadpole cancellation
condition can be interpreted as a charge conservation requirement,
rather than a static constraint.

In \cite{stw,acfi} another set of ``flux'' deformations was
considered in the effective theory in order to restore mirror
symmetry between type IIB and type IIA orientifold
compactifications in the presence of non-vanishing 3-form fluxes.
The idea is that T-duality acts on the NSNS 3-form flux through
the chain
\begin{equation}
H_{mnp}{\stackrel{\T_m}
{\longleftrightarrow}}-\omega^m_{np}{\stackrel{\T_n}
{\longleftrightarrow}}-Q^{mn}_p{\stackrel{\T_p}
{\longleftrightarrow}}R^{mnp}\ .
\end{equation}
Since T-duality exchanges Kaluza-Klein and winding modes, the
stringy nature of these deformations increases as we move to the
right in this chain and, in particular, ten dimensional
supergravity starts failing to be a good description. Thus,
$\omega$-fluxes are geometric fluxes, describing compactifications
on twisted tori with structure constants $\omega^i_{jk}$, as
discussed below. Q-fluxes parameterize compactifications on
T-folds, where T-duality is used, together with diffeomorphisms,
to glue different local patches of the manifold together. These
are globally non-geometric constructions, but locally geometric,
sometimes related to the effective ``geometry'' of asymmetric
orbifolds \cite{asim1,asim2,asim3,mcgreevy}. Finally, R-fluxes
correspond to highly non-geometric compactifications, which do not
even admit a local geometric description.

The orientifold projection  we have chosen projects  all
the geometric and R-fluxes out. Only,  24 components of
$Q^{ij}_k$ survive the projection, as summarized in the Appendix.
These can be thought as the structure constants of a gauge algebra
\cite{stw,acfi}
\begin{equation}
[X^i,X^j]=Q^{ij}_pX^p\ ,\label{xxq}
\end{equation}
where $X^i$ are the gauge group generators associated to vector
fields which, in the T-dual type I picture, result from
dimensional reduction of the metric, as we will see in a moment.
In particular, Q-fluxes must satisfy the Jacobi identity of the
algebra
\begin{equation}
Q^{[mn}_pQ^{r]p}_t=0\ .\label{qq}
\end{equation}
To gain a more intuitive understanding of this algebraic
structure, a brief parenthesis  describing the  T-dual type I
compactification with O9-planes is worthwhile. Performing six
T-dualities along the internal 6-torus, Q-fluxes are mapped into
geometric fluxes
\begin{equation}
Q_p^{qr}\to\omega^p_{qr}\ .
\end{equation}
These  have a clear interpretation from the point of view of the
compactified theory.  In an ordinary Kaluza-Klein compactification
on a torus, the different fields only depend on the space-time
coordinates and we must deal with exact differentials $dx^i$ for
the internal tori. However,  a Scherk-Schwarz \cite{Scherk:1979zr}
like compactification can be achieved by allowing for a controlled
dependence of the fields on the internal coordinates $x^a$
($a=1,\dots 6$).  The reduction is realized in terms of an
expansion around a non-trivial metric described by new twisted
tori vielbeins $\eta^c= U^c_ p(x) \, dx^p$. The functions $ U^c_
p(x)$, encoding  the dependence on internal coordinates, are not
completely arbitrary. They must be such that the vielbein is a
globally defined basis of 1-forms such that
\beq d\eta^c = -\oh \omega^c_{ab} \eta^a \wedge \eta^b \ ,
\label{gflux} \eeq
with $ \omega^c_{ab}$ constant.

For the vector bosons, $A_{\mu}^a \equiv g_{\mu}{}^a$, built up
from the ten dimensional metric $g_{mn}$, the gauge
transformations read \cite{DallAgataferrara}
\begin{equation}
\delta g_{\mu}{}^c= \partial_{\mu}\lambda^c-\omega_{ab}^c\lambda^a
g_{\mu}{}^b\ ,
\end{equation}
where $\lambda^c$ are the group parameters. Thus, the metric
fluxes $\omega_{ab}^c$ correspond to the gauge group structure
constants. By introducing the gauge group generators $Z_p$
\begin{equation}
A_{\mu}=g_{\mu}{}^p Z_p\ ,
\end{equation}
we find
\begin{equation}
[Z_a, Z_b]  = {\omega}_{ab}^p Z_p \label{zzi}
\end{equation}
with $a=1,\dots 6$. This is not other but the T-dual algebra to
eq.(\ref{xxq}). The Jacobi identities of the algebra, for each
$Z_l$, lead to the constraints,
\begin{eqnarray}
  {\omega}_{[ab}^p\,{\omega}_{c]p}^l & = & 0\ ,
\label{ww}
\end{eqnarray}
which correspond to the Bianchi identities derived from
(\ref{gflux}) and which are dual of (\ref{qq}). In Section
\ref{sec4} we will find convenient to use this type I description
to elaborate on the Freed-Witten constraints which arise from the
flux algebra.

Let us now  return to the type IIB picture with Q-fluxes. From the
ten dimensional point of view, the $Q^{ij}_k$ fluxes lead to local
supergravity solutions which interpolate between different
standard $\mathcal{N}=1$ orientifold involutions \cite{fer} (see
also \cite{mariana,dall}). Qualitatively, these are characterized
by a non-constant holomorphic axion-dilaton which transforms under
global monodromies of the manifold, with $S(x^k+1)$ being related
to $S(x^k)$ through T-duality transformations.

In many senses, the background locally resembles a standard type
IIB compactification with 3-form fluxes. However, the B-field
contains an extra dependence on the dilaton,
\begin{equation}
B_2\simeq e^{\phi(x^k)}\ Q^{ab}_k x^k dx^a\wedge dx^b
+\ldots\label{linearb}
\end{equation}
Thus, due to the term in the Wess-Zumino action
\begin{equation}
\int C_6\wedge B_2\ ,
\end{equation}
a D7-brane wrapping a 4-cycle which contains the 2-cycle
$[dx^a\wedge dx^b]$, experiences a solitonic D5-brane charge in
the worldvolume which, due to the non-linear dependence of the
dilaton on $x^k$, cannot be absorbed by means of a gauge
transformation of the B-field. In order to perform the
identification $x^k\sim x^k+1$, a stringy monodromy (two T-duality
transformations along $x^a$ and $x^b$) is required, so that the
change in the dilaton is compensated.

In the four dimensional theory, the deformation on the moduli
space induced by these stringy monodromies can be parameterized in
terms of the superpotential \cite{acfi}
\begin{equation}
W=\int(F_3-iSH_3+Q\mathcal{J}_c)\wedge \Omega\ ,\label{qsuper}
\end{equation}
where we have defined the $Q$-product as
\begin{equation}
(QX)_{pm_1\ldots m_{p-2}}=\frac12 Q^{ab}_{[p}X_{m_1\ldots
m_{p-2}]ab}\ .
\label{QX}
\end{equation}
In addition, Q-fluxes induce a D7-brane charge
\begin{equation}
-\int C_8\wedge QF_3\ ,\label{qtadpole}
\end{equation}
which must be cancelled by the effect of localized sources
(D7-branes and/or O7-planes). By analogy with the D3-brane case,
it is then natural to conjecture that Q-fluxes are related to
D7-branes through (non-)geometric transitions.

At first sight, it may seem surprising that the four dimensional
effective theory captures part of the stringy nature of the
original compactification. Indeed, it is perhaps more appropriate
to think of the four dimensional effective supergravity as being
the low energy limit of a ten dimensional string setup, rather
than a ten dimensional supergravity background. States of the
higher dimensional theory which are protected by holomorphicity
are expected to be captured by the superpotential in the four
dimensional effective supergravity. Up to what extent the
superpotential (\ref{qsuper}) gives a complete description of the
light modes in compactifications with non-vanishing Q-fluxes is,
however, still an open question.

\subsection{Non-geometric compactification of F-theory}
\label{secf}

Type IIB string theory is conjectured to be invariant under
S-duality. This is reflected in a $SL(2,\mathbb{R})$ self-duality
of the type IIB supergravity equations of motion, broken to
$SL(2,\mathbb{Z})$ at the quantum level. Moreover, whereas in ten
dimensions S-duality is a weak-strong coupling duality, leading to
the exchange of winding modes and E1-instanton effects, in four
dimensions it manifests as weak-weak electric-magnetic duality.
This is the idea behind the adiabatic argument of Vafa-Witten
\cite{adiabatic}, providing us with a prescription for building
S-dual pairs of ten dimensional superstring compactifications. It
is therefore natural to conjecture that the dynamics of some light
modes, protected by holomorphicity, in strongly coupled
compactifications of type IIB string theory, is partially captured
by gaugings in the four dimensional effective supergravity.

The complex axion-dilaton transforms under $SL(2,\mathbb{Z})$ as
\begin{equation}
S\to \frac{kS-i\ell}{imS+n}\ , \quad kn-\ell m = 1\ , \quad k,\
\ell,\ m,\ n\in \mathbb{Z}\ ,
\end{equation}
whereas 3-form fluxes behave as a doublet
\begin{equation}
\begin{pmatrix}F_3\\ H_3\end{pmatrix}\to\begin{pmatrix}k&\ell\\
m&n\end{pmatrix}\begin{pmatrix}F_3\\ H_3\end{pmatrix},
\end{equation}
and $C_4$ remains invariant. With these transformation rules at
hand, we may easily check that the D3-brane tadpole cancellation
condition, eq.(\ref{d3tadpole}), and the flux superpotential,
eq.(\ref{gvw}), are invariant under S-duality, up to K\"ahler
transformations of the superpotential.

When D7-branes and/or $Q$-fluxes are also present, the
$SL(2,\mathbb{Z})$ invariance of the theory becomes more subtle.
The RR 8-form, $C_8$, is one of the components of a
$SL(2,\mathbb{Z})$ triplet of 8-forms, $(C_8,\tilde C_8, C'_8)$
\cite{7s1,7s2,7s3,bergshoef}. Under $S\to 1/S$ the elements of the
triplet transform as
\begin{align}
C_8&\to -\tilde C_8\ , \\
\tilde C_8&\to -C_8\ , \\
C'_8&\to -C'_8\ .
\end{align}
There are 7-branes coupling electrically to all these forms, with
charges that we denote as $p^2$, $q^2$ and $r$, respectively.
These can be arranged into a $SL(2,\mathbb{Z})$-invariant
traceless matrix of charges
\begin{equation}
\mathcal{N}_7\equiv \begin{pmatrix}r/2&p^2\\
-q^2&-r/2\end{pmatrix}\ . \label{n7}
\end{equation}
The charge matrix $\mathcal{N}_7$ determines the
$SL(2,\mathbb{Z})$ monodromy suffered by the axion-dilaton when
going in a closed path around the 7-brane \cite{bergshoef}
\begin{equation}
S\to
\left[\textrm{cos}(n_7^{1/2})I+\frac{\textrm{sin}(n_7^{1/2})}{n_7^{1/2}}\mathcal{N}_7\right]S\
,
\end{equation}
with $n_7\equiv \textrm{det }\mathcal{N}_7$ and $I$ the $2\times
2$ identity matrix.

Physical 7-branes transform as non-linear doublets of
$SL(2,\mathbb{Z})$ \cite{bergshoef}, so that there are only two
independent entries in $\mathcal{N}_7$. Different
$SL(2,\mathbb{Z})$ orbits are characterized by different values of
$n_7$.

In the uplift to F-theory, $(p,q)$ 7-branes are associated to
codimension two loci where the elliptic fiber, which parameterizes
the axion-dilaton dependence on the compact coordinates,
degenerates into an ADE singularity. Therefore, F-theory $(p,q)$
7-branes lay in the same orbit as the D7-brane, characterized by
the condition $n_7=0$ \cite{bergshoef}.

We denote the total charge associated to $(p,q)$ $7_k$-branes by
the matrix
\begin{equation}
(\mathcal{N}_{7_k})_{\rm total}=\sum_I \begin{pmatrix}p^I_kq^I_k&(p^I_k)^2\\
-(q^I_k)^2&-p^I_kq^I_k\end{pmatrix}\ ,\label{7total}
\end{equation}
where the sum extends over all $7_k$-branes present in the model.
In general, det $\left(\mathcal{N}_{7_k}\right)_{\rm total}\neq
0$, and the configuration cannot be rotated by a
$SL(2,\mathbb{Z})$ transformation into a setup with only
$D7$-branes. The configuration is, thus, intrinsically
non-perturbative. On the other hand, objects with $n_7\neq 0$ can
be seen as non-perturbative bound states of $(p,q)$ 7-branes.
Notice that O7-planes are also implicitly included in
(\ref{7total}), as they are the weak coupling limit of compound
objects made up of several non-perturbative $(p,q)$ 7-branes
\cite{sen,bergshoef}.

From eqs.(\ref{qsuper}) and (\ref{qtadpole}) it is evident that
Q-fluxes must transform in a non-trivial way to keep the
$SL(2,\mathbb{Z})$ invariance of the theory. In \cite{acfi}, it
was conjectured that such fluxes  belong to a linear doublet of
$SL(2,\mathbb{Z})$, so a new set of flux parameters, $P^{ij}_k$,
was considered
\begin{equation}
\begin{pmatrix}Q\\ P\end{pmatrix}\to\begin{pmatrix}k&\ell\\
m&n\end{pmatrix}\begin{pmatrix}Q\\ P\end{pmatrix}\ ,
\end{equation}
We summarize in the Appendix the 24 components of $P^{ij}_k$ which
are compatible with the $\mathbb{Z}_2\times \mathbb{Z}_2$
symmetry.

The effective superpotential and Chern-Simons couplings,
eqs.(\ref{qsuper}) and (\ref{qtadpole}), must  be extended
accordingly to the expressions
\begin{equation}
W=\int [(F_3-iSH_3)+(Q-iSP)\mathcal{J}_c]\wedge\Omega\
.\label{pqsuper}
\end{equation}
and
\begin{equation}
\int -C_8\wedge QF_3+\tilde C_8\wedge PH_3+C'_8\wedge(QH_3+PF_3)\
,
\label{c8couplings}
\end{equation}
Therefore, $P$ and $Q$-fluxes naturally source the tadpole
cancellation requirements associated to the total $(p,q)$
$7_k$-brane charge
\begin{align}
&(QH_3+PF_3)_k=2\sum_Ip_k^Iq_k^I\ ,\label{7can}\\
&(QF_3)_k=\sum_I (p_k^I)^2\ ,\label{d7can}\\
&(PH_3)_k=\sum_I (q_k^I)^2\ ,\label{ns7can}
\end{align}
where we have defined the P-product in a similar way to the
Q-product, eq.(\ref{QX}), and we have expanded the l.h.s. in a
basis of 2-forms, $\omega_k$, $k=1,2,3$, dual to $\tilde
\omega_k$.

In a similar fashion to what occurs with the rest of the flux
parameters, it is natural to expect that P-fluxes also must
satisfy a set of Bianchi identities. The ansatz for some of these
constraints was already presented in \cite{acfi}, based on
symmetry arguments. In Section \ref{sec2}, however, we derive in a
more systematic and precise way the full set of constraints that
the P-fluxes must satisfy, finding some differences with
respect to \cite{acfi}.

Since the $SL(2,\mathbb{Z})$ symmetry, relating P and Q-fluxes, is
an explicit symmetry of the type IIB supergravity equations of
motion, we also expect the $P$-flux parameters to correspond to
compactifications which admit a ten dimensional local supergravity
description. Indeed, acting with $SL(2,\mathbb{Z})$ in each local
patch of a compactification with non-vanishing Q-flux, we observe
that (\ref{linearb}) is translated into a dilaton dependence for
the RR 2-form potential, whenever a non-vanishing $P^{ij}_k$ flux
is turned on
\begin{equation}
C_2\simeq e^{\phi(x^k)}P^{ab}_k x^k dx^a\wedge dx^b +\ldots\ .
\end{equation}
Hence, for instance, a $(0,1)$ 7-brane wrapping a 4-cycle
containing $[dx^a\wedge dx^b]$, will experience a solitonic
NS5-brane charge depending on its position in $x^k$, induced by
the coupling in the Wess-Zumino action
\begin{equation}
\int B_6 \wedge C_2\ .
\end{equation}
The stringy monodromy gluing $x^k\simeq x^k+1$, in this case
involves an S-duality transformation followed by two T-dualities
along $x^a$ and $x^b$ and a further S-duality. Since S and
T-dualities do not commute, the total monodromy is not equivalent
to just performing two T-dualities.

For vanishing fluxes, holomorphicity and neutrality under
monodromies are enough to determine the profile of $S$ in terms of
the modular invariant holomorphic $j$-function \cite{sen}. In the
presence of Q and/or P-fluxes, however, the holomorphic
axion-dilaton, $S$, is no longer monodromy neutral. In the next
section, we present some simple explicit examples which will serve
to clarify the above discussion.

\subsection{A non-geometric limit for F-theory on K3}
\label{secfth}

Let us consider Sen's construction of F-theory compactified on a
elliptically fibered K3 \cite{sen}. The later is given by the
Weierstrass curve
\begin{equation}
\hat y^2=\hat x^3+P(z)\hat x+Q(z)\ ,
\end{equation}
where $z=x^3+iU_3x^6$ is the coordinate of the $CP^1$ base, $\hat
x$ and $\hat y$ the coordinates of the fibre, and $P^3+Q^2$ is a
polynomial of order 24, whose zeros correspond to the locations of
the 7-branes transverse to the $CP^1$. Global consistency requires
24 D7-branes. The axion-dilaton is given in terms of the modular
parameter of the fibre
\begin{equation}
S(z)=-ij^{-1}\left(\frac{P^3(z)}{P^3(z)+Q^2(z)}\right)\ ,
\end{equation}
with $j$ the modular invariant $j$-function. In the limit on which
$P^3(z)\propto Q^2(z)$, the axion-dilaton becomes constant and the
$CP^1$ degenerates into a $T^2/\mathbb{Z}_2$ orbifold. Each of the
four singularities,
$(x^3,x^6)=\{(0,0),(1/2,0),(0,1/2),(1/2,1/2)\}$, corresponds to a
bound state of $(p,q)$ 7-branes, which in the weak coupling limit
manifests as 4 D7-branes, each one with charge +1, and a single
O7-plane, with charge -4. The four dimensional theory resulting
from further compactification in $T^2_{[x^1x^4]}\times
T^2_{[x^2x^5]}$ is $\mathcal{N}=4$ supersymmetric. More
generically, some of the 7-branes may be magnetized, leading to
$\mathcal{N}=2,1,0$ supersymmetry in four dimensions.

Being slightly more general, we can perform an $SL(2,\mathbb{Z})$
rotation of the above system by the matrix
\begin{equation}
\Lambda_{(\mathfrak{p},\mathfrak{q})}=\begin{pmatrix}\mathfrak{p}&\mathfrak{a}\\
-\mathfrak{q}&\mathfrak{b}\end{pmatrix}\ ,\label{sl}
\end{equation}
with $(\mathfrak{p},\mathfrak{q})$ two mutually prime integers,
and $(\mathfrak{a},\mathfrak{b})$ one of the infinite solutions of
the B\'ezout's identity
\begin{equation}
\mathfrak{p} \mathfrak{b}+\mathfrak{q} \mathfrak{a}=1\ .
\end{equation}
After performing this rotation, the 16 D7-branes become 16
$(\mathfrak{p},\mathfrak{q})$ 7-branes, whereas each of the
O7-planes becomes a different bound state of $(p,q)$ 7-branes,
with total charge $-4(\mathfrak{p},\mathfrak{q})$.

Based on the tadpole cancellation conditions,
eqs.(\ref{7can})-(\ref{ns7can}), we expect $(p,q)$ 7-branes to be
related to Q and P-fluxes by non-geometric transitions. We would
like to consider here the case on which $N^2/4$ of the 4
$(\mathfrak{p},\mathfrak{q})$ 7-branes at each singularity, with
$N=0,2,4$, have been traded by fluxes. The resulting closed string
background must satisfy the 7-brane charge conservation
conditions, given by eqs.(\ref{7can})-(\ref{ns7can}). A possible
choice of the flux parameters satisfying these requirements is
given by,
\begin{align}
Q^{42}_6=-Q^{15}_6=-F_{423}=F_{153}=\mathfrak{p}N \ , \label{back}\\
P^{15}_6=-P^{42}_6=-H_{153}=H_{423}=\mathfrak{q} N \ .\nonumber
\end{align}
Indeed, plugging (\ref{back}) into (\ref{pqsuper}) it is not
difficult to check that there is a $\mathcal{N}=2$ supersymmetric
Minkowski vacuum with $U_3=T_3$ and $U_1=U_2$. Interestingly
enough, performing an inverse $SL(2,\mathbb{Z})$ transformation,
$\Lambda_{(\mathfrak{p},\mathfrak{q})}^{-1}$, followed by 4
T-dualities along $T^2_{[x^1x^4]}\times T^2_{[x^2x^5]}$, the
background (\ref{back}) is dual to a type IIB toroidal
compactification with standard 3-form fluxes, whose ten
dimensional solution is well known \cite{dasg,gp,gkp}. Hence, in
the same spirit than \cite{fer}, we can perform the above chain of
dualities, starting with the dual background, to obtain the ten
dimensional supergravity solution associated to the flux
parameters (\ref{back}).

The initial type IIB configuration with 3-form fluxes is the same
than the one considered in \cite{fer}. Let us concentrate first in
the case where $\Lambda_{(\mathfrak{p},\mathfrak{q})}=\mathbb{I}$.
Performing 4 T-dualities and a redefinition of the moduli
parameters, the resulting ten dimensional solution can be written
as\footnote{We thank F. Marchesano for sharing with us some of his
notes where this setup was also considered.},
\begin{align}
ds^2&=Z^{-1/2}ds^2_{\mathbb{R}^{1,3}}
+Z^{1/2}\left[\sum_{k=1,2}\frac{t_3e^{\phi}t_k}{\textrm{Re
}U}|dx^k+iUdx^{k+3}|^2+\frac{Z^{1/2}}{g_s
t_3}|dx^3+iT_3dx^{6}|^2\right]\ , \nonumber \\ C_2&=N
x^3(dx^1\wedge dx^5-dx^4\wedge dx^2)\ , \label{c2}\\ B_2&=N
e^{\phi}t_3x^6(dx^1\wedge dx^5-dx^4\wedge dx^2)\ ,
\\
F_1&=*_{T^2_{[x^3x^6]}}de^{-\phi}\ ,\label{f1}\\
e^\phi&=\frac{t_3^{-1}}{(Nx^6)^2+t_1t_2 Z}\ ,\label{di}\\
F_5&=C_2\wedge H_3+F_3\wedge B_2-\frac12 F_1\wedge B_2\wedge
B_2-C_0\wedge H_3\wedge B_2\ ,\nonumber
\end{align}
where $U$, $g_s$, $t_{1,2}$ and $t_3\equiv \textrm{Re }T_3$ are
constant modular parameters, and $Z(x^3,x^6)$, with is the warp
factor in the dual type IIB configuration with 3-form fluxes. In
terms of these parameters, $Z(x^3,x^6)$ satisfies the equation
\begin{equation}
-\tilde
\nabla^2_{T^2_{[x^3x^6]}}Z=\frac{2g_sN^2}{t_1t_2t_3}-\frac{N^2}{2}\sum_{m,n=0,1/2}\delta(x^3-m,x^6-n)
\end{equation}
From eqs.(\ref{f1}) and (\ref{di}), we observe that in order to
compute $S$ we have to solve this equation. From an open string
dual perspective, this is equivalent to compute the gauge
threshold corrections of a probe D3-brane \cite{liam}.

Making use of the results of \cite{liam} (see also \cite{bhk}) we
obtain
\begin{equation}
Z=1-\frac{N^2}{t_1t_2t_3}\left[\frac{(\textrm{Im
}z)^2}{t_3}-\frac{1}{8\pi}\left|\vartheta_1(2z,-iT_3)\right|^2\right]\
,
\end{equation}
which plugged into (\ref{c2})-(\ref{di}) results in,
\begin{align}
S&=t_1t_2t_3+\frac{N^2}{8\pi}\textrm{log }\vartheta_1(2z,-iT_3)\ ,\label{s0}\\
C_2&=N x^3(dx^1\wedge dx^5-dx^4\wedge
dx^2)\ , \\
B_2&=\frac{Nt_3x^6}{t_1t_2t_3+\frac{N^2}{8\pi}\textrm{log
}\left|\vartheta_1(2z,-iT_3)\right|^2}(dx^1\wedge dx^5-dx^4\wedge
dx^2)\ .\label{b2}
\end{align}
The moduli space of the resulting non-geometric compactification
is given by the positions of the $24-N^2$ D7-branes. Indeed, under
four T-dualities these are mapped to D3-branes, which do not
receive supersymmetric mass terms from 3-form fluxes, and
therefore can move freely on the bulk.

When one of the D7-branes at the origin moves along the $x^6$
direction, it develops a charge of D5-brane in its worldvolume,
induced by the Chern-Simons coupling of $C_6$ to $B_2$, the latter
given in eq.(\ref{b2}), in the worldvolume action. In order to
identify $x^6\simeq x^6+1$, two T-dualities are needed in the
corresponding transition function, as pointed out in \cite{fer}.

Compactifications with non-vanishing Q-flux are often related, in
special points of the moduli space, to asymmetric orbifolds
\cite{asim1,asim2,asim3,mcgreevy}. It is tempting to believe that,
in the case at hand, where some of the D7-branes in the
$T^4/\mathbb{Z}_2$ orbifold limit of F-theory on a K3 have been
traded by fluxes, there is still a valid (asymmetric) orbifold
description. If such is the case, the geometric moduli of the
remaining D7-branes would parameterize the corresponding
``blow-up'' modes of the asymmetric orbifold. However, wether this
is a valid description remains an open question.

To finish this section, let us consider now the general case, in
which $\Lambda_{(\mathfrak{p},\mathfrak{q})}$ is an arbitrary
$SL(2,\mathbb{Z})$ rotation given in eq.(\ref{sl}). It is simple
to see that eqs.(\ref{s0})-(\ref{b2}) are modified to
\begin{align}
S&= \frac{8\pi(\mathfrak{p} t_1t_2t_3-i\mathfrak{a})+\mathfrak{p}
N^2\textrm{log
}\vartheta_1(2z,-iT_3)}{8\pi(\mathfrak{b}-i\mathfrak{q}
)-i\mathfrak{q} N^2 \textrm{log }\vartheta_1(2z,-iT_3)}\
, \label{sss}\\
 C_2&=N\left(\mathfrak{p}
x^3+\mathfrak{a}\frac{t_3x^6}{t_1t_2t_3+\frac{N^2}{8\pi}\textrm{log
}\left|\vartheta_1(2z,-iT_3)\right|^2}\right)(dx^1\wedge
dx^5-dx^4\wedge dx^2)\ , \\
B_2&=N\left(-\mathfrak{q}
x^3+\mathfrak{b}\frac{t_3x^6}{t_1t_2t_3+\frac{N^2}{8\pi}\textrm{log
}\left|\vartheta_1(2z,-iT_3)\right|^2}\right)(dx^1\wedge
dx^5-dx^4\wedge dx^2)\ ,
\end{align}
corresponding to the ten dimensional local realization of the flux
parameters (\ref{back}), whereas the metric and the RR 5-form are
independent of $\Lambda_{(\mathfrak{p},\mathfrak{q})}$. Now there
are $4-(N/2)^2$ $(\mathfrak{p},\mathfrak{q})$ 7-branes at each of
the four singularities, which can move freely along
$T^2_{[x^3x^6]}$. In this case, the Wess-Zumino action for $(p,q)$
7-branes contains a term \cite{7inv}
\begin{equation}
p\ C_6\wedge B_2+q\ B_6\wedge C_2+\ldots
\end{equation}
so that $(\mathfrak{p},\mathfrak{q})$ 7-branes develop a
$(\mathfrak{p},\mathfrak{q})$ 5-brane charge in its worldvolume
depending on their position in $x^6$ and an $SL(2,\mathbb{Z})$
rotation by $\Lambda_{(\mathfrak{p},\mathfrak{q})}^{-1}$, followed
by two T-dualities and an inverse rotation by
$\Lambda_{(\mathfrak{p},\mathfrak{q})}$, have to be carried on in
the transition functions gluing different patches, as it was
advanced in the previous section.

In general, the axion-dilaton $S$, will also depend on the
positions of the $16-N^2$ 7-branes, as occurs in the flux-less
case. Here, we have solved for the warping in the special point of
the moduli space where the 7-branes are equally distributed on top
of the four singularities. The expression we have found for the
axion-dilaton, eq.(\ref{sss}), in general is not valid away from
this special point. However, although $C_2$ and $B_2$ depend also
on $S$ we expected that, to leading order, the dependence of these
fields on $x^6$ is always $\mathcal{O}(x^6)$.

%%%%%%%%%%%%%%%%%%%%%%%%%%%%%%%%%%%%%%%%%%%%%%%%%%%%%%%%%%%%%%%%%%%%%%%%%%%%%%

\section{Flux algebra from  $SL(2,\Z)^7$ dualities}
\label{sec2}

\subsection{Flux algebra for $F_3$ and Q-fluxes}
\label{sec20}

We proceed now to derive the full gauge algebra, starting from
eq.(\ref{xxq}), associated to Kaluza-Klein vectors in the class of
non-geometric F-theory compactifications described in the previous
section, and see what kind of information we can extract.

The global symmetry group associated to the moduli space of the
factorized 6-torus is $SL(2,\mathbb{Z})^7$, where each of the
$SL(2,\mathbb{Z})_k$ factors acts on a different modulus and
consists of the generators
\begin{equation}
\mathcal{S}_{k,1}=\begin{pmatrix}1&1\\ 0&1\end{pmatrix}\ , \qquad
\mathcal{S}_{k,2}=\begin{pmatrix}0&1\\ -1&0\end{pmatrix}\ , \qquad
k=0,\ldots , 6.
\end{equation}
The seven untwisted moduli can be grouped in the vector
representation of $SL(2,\mathbb{Z})^7$
\begin{equation}
\mathbb{T}\equiv \mathbf{7}=(S;\ T_1,\ T_2,\ T_3;\ U_1,\ U_2,\
U_3)\ ,
\end{equation}
whereas the flux parameters surviving the $\mathbb{Z}_2\times
\mathbb{Z}_2$ projection transform in the spinorial
representation, $\mathbb{G}\equiv\mathbf{128}$ \cite{acfi}.
Labelling the components of $\mathbb{G}$ by the set of weights
$(\pm,\pm,\pm,\pm,\pm,\pm,\pm)$, where $\pm$ stands for
$\pm\frac12$, under $\mathbb{T}_k\to 1/\mathbb{T}_k$ the flux
parameters transform as
\begin{equation}
\mathcal{S}_{k,2}(n_0,\ldots,n_k,\ldots,n_6)=\textrm{sign}(n_k)(n_0,\ldots,-n_k,\ldots,n_6)\
.
\end{equation}
Notice that 64 of these flux parameters correspond to the $H_3$,
$F_3$, $Q$ and $P$-fluxes discussed in the previous section,
describing F-theory compactifications on U-folds.

There is a $SL(2,\mathbb{Z})^7$ invariant (up to K\"ahler
transformations) superpotential induced by the flux spinor
$\mathbb{G}$, which is given by \cite{acfi}
\begin{equation}
W=\left. \mathbb{G}\otimes
e^{i\mathbb{T}}\right|_{(+;+,+,+;+,+,+)}\label{super}
\end{equation}
with $(+;+,+,+;+,+,+)$ the spinor component selected by the four
dimensional $\mathcal{N}=1$ gravitino, and
\begin{equation}
e^{i\mathbb{T}}\equiv
1+i\mathbb{T}-\mathbb{T}\otimes\mathbb{T}+\ldots
\end{equation}
The expression of the superpotential in terms of flux parameters
can be found in the Appendix.

In terms of $SL(2,\Z)^7$ Weyl spinors, the 24 components of the
Q-flux allowed by the orientifold involution read
\begin{equation}
Q^{ab}_p \equiv (+;{\underline {-,+,+}};{\underline {\pm,\pm,\pm}})\\
\end{equation}
where underlining indicates all possible permutations.

Since we know how $SL(2,\Z)^7$ transformations act on the fluxes,
a systematic  use of such actions on eq.(\ref{xxq}) should lead to
the complete algebra involving the  $2^7$ dual fluxes.

With this purpose in mind, we assign spinor indices to the $X^a$
generators. By matching the left and right sides of eq.(\ref{xxq})
we get\footnote{This is the index structure that would be assigned
to $X^a\equiv {(Q^a )}^b_c$ by identifying the structure constants
as matrix elements of generators in the adjoint representation.
For instance, $X^1={(Q^1) }^b_c$ corresponds to a $6 \times 6$
matrix with eight non-vanishing elements given by fixing a plus
sign in the $0^{\rm th}$, $1^{\rm st}$ and $4^{\rm th}$ components
of the weight vector and keeping the other components unfixed (see
table \ref{mapa} in the Appendix).}
\begin{equation}
X^i
={(+;{\stackrel{\mathsmaller{i}}{\overbrace{+,0,0}};\stackrel{\mathsmaller{i+3}}{\overbrace{+,0,0}}})}\
,\qquad X^{i+3} =
{(+;{\stackrel{\mathsmaller{i}}{\overbrace{+,0,0}};\stackrel{\mathsmaller{i+3}}{\overbrace{-,0,0}}})}\
,
\end{equation}
where a ``zero'' in a given position reflects that the generator
behaves as a scalar under the action of that particular
$SL(2,\mathbb{Z})$ .

Now, it is simple to see that the algebra (\ref{xxq}) is
incomplete. Consider for instance the following commutator in
eq.(\ref{xxq}),
\begin{multline}
[X^1, X^2] = Q^{12}_3 X^3+Q^{12}_6 X^6 = \\
  \ = \ -(+;+,+,-;+,+,-)X^3\ +\ (+;+,+,-;+,+,+)X^6\ .
\label{z1z2}
\end{multline}
The l.h.s. in this equation is invariant under
$SL(2,\mathbb{Z})_k$, for  $k=3,6$, whereas the r.h.s. is not.
Hence, if we insist in consistency under those generators we must
modify the right member of the equality.

Acting with $\mathcal{S}_{3,2}$ on the fluxes
\begin{equation}
\mathcal{S}_{3,2}Q^{12}_3=-\tilde{F}^{126} \ , \qquad
\mathcal{S}_{3,2}Q^{12}_6=\tilde{F}^{123} \ ,
\end{equation}
with $\tilde{F}_3$ defined in eq.(\ref{tildeF}), and introducing a
new set of generators, $Z_i$
\begin{equation} Z_i
={-(+;{\stackrel{\mathsmaller{i}}{\overbrace{-,0,0}};\stackrel{\mathsmaller{i+3}}{\overbrace{-,0,0}}})}\
,\qquad Z_{i+3} =
{(+;{\stackrel{\mathsmaller{i}}{\overbrace{-,0,0}};\stackrel{\mathsmaller{i+3}}{\overbrace{+,0,0}}})}\
, \end{equation}
the modified  invariant commutator is given by,
\begin{eqnarray}
[X^1, X^2] & = & Q^{12}_3 X^3+Q^{12}_6 X^6-{\tilde
F}^{123}Z_3-{\tilde  F}^{126}Z_6 \ .\label{x1x2}
\end{eqnarray}
The application of ${\cal S}_{6,2}$ would lead to the same
conclusion. Actually, we could have envisaged this expression by
recalling that a $SL(2,\Z)$ scalar built up from two spinors
$W_\pm$ and $Q_\pm$ takes the expression,
\begin{equation}
 W_+ Q_{-}-W_- Q_+ \equiv W_p Q^p\ .
\end{equation}
While the sixth index in eq.(\ref{z1z2}) is contracted in an
invariant way, the third index is not.
For the generic case we therefore have the algebra,
\begin{eqnarray}
[X^a, X^b] & = &- \, {\tilde F}^{abp} Z_{p}+Q^{ab}_p X^p\ .
\label{zazb}
\end{eqnarray}

The same kind of reasoning allows to compute the other
commutators, $[Z_a,X^b]$ and $[Z_a,Z_b]$. For example, applying
${\cal S}_{b,2}$ to the commutator $[X^a,X^b]$ and using the rules
derived above, we obtain  $[X^a,Z_{b+3}]$.
Thus, by proceeding in the same manner with the other generators,
we find a full algebra involving non-geometric and RR 3-form
fluxes,\footnote{In this work we only consider $F_3$, $H_3$, $Q$
and $P$-fluxes, as discussed in Section \ref{sec1}. Strictly
speaking, the above procedure would lead also to the inclusion of
additional sets of \emph{primed} fluxes, $F'_{abc}$ and
$\omega'^a_{bc}$, so that the complete invariant algebra is,
\begin{align*}
[X^a, X^b] & = - \ {\tilde  F}^{abp} Z_{p}+{Q}^{ab}_p X^p\ , \\
{[X^a, Z_b]} & =  \ {\omega'}_{ap}^{b} X^{p}-{Q}^{ap}_b Z_p\ , \\
{[Z_a, Z_b]} & =  \ -F'_{abp} X^{p}+{\omega'}^{p}_{ab} Z_p\ .
\end{align*}
\label{pri}}
\begin{eqnarray}
[X^a, X^b] & = & -\ {\tilde F}^{abp} Z_{p}+{Q}^{ab}_p X^p\ , \label{fqalgebra0}\\
{[X^a, Z_b]} & = &  -{Q}^{ap}_b Z_p\ , \nonumber\\
{[Z_a, Z_b]} & = &0\ . \nonumber
\end{eqnarray}
The Jacobi identity of this algebra is given by eq.(\ref{qq})
together with the additional constraint,
\begin{equation}
{Q}^{[ab}_p \, {\tilde F} ^{c]lp}\ + \ \tilde F^{p[ab} \,
{Q}^{c]l}_p\ = \ 0 \quad \Leftrightarrow \quad QF_3=0\ ,
\label{qf}
\end{equation}
which can be identified with the D7-brane flux-induced charge
cancellation condition, eq.(\ref{d7can}), with $\sum_I (p_k^I)^2=0$.
The fact that the tadpole cancellation conditions arising from
this algebra do not contain contributions from localized sources,
is consistent with the missing additional generators associated to
extra massless vector multiplets. The latter are required to
account for the gauge symmetries of the field theories confined to
the worldvolume of the branes. In Section \ref{sec4} we give a
detailed discussion on how to extend the algebras of closed string
gauge generators, to account also for the open
string\footnote{With some abuse of language, we generically refer
to  the  ``open string sector''  to describe the sector which
contains the fields on the worldvolume of the branes even if in
the generic cases of $(p,q)$ 7 branes such fields  are not
associated to fundamental open strings. We thank A. Uranga for a
comment on this point. } massless vector fields.

The algebra (\ref{fqalgebra0}) is consistent with other results
that have previously appeared in the literature. Thus, for
example, its type I T-dual version,
\begin{eqnarray}
[Z_a, Z_b] & = &- \ {\bF}_{abp} Y^{p}+{\omega}_{ab}^p Z_p\ ,\\
{[Z_a, Y^b]} & = & -{\omega}_{ap}^b Y^p\ ,\nonumber\\
{[Y^a, Y^b]} & = & 0\ ,\nonumber
\end{eqnarray}
where we have made use of Tables \ref{rriib} and \ref{nongeoiib}
in the Appendix and we have introduced $Z_a$ and $Y^a$ as the
T-duals of $X^a$ and $Z_a$, respectively, is the algebra found in
\cite{DallAgataferrara}, in the framework of supergravity
compactifications on twisted tori. In that context, $ Y^b$  are
the generators associated to the RR vector field, $ C_{\mu}=C
_{\mu p} Y^p$, obtained by dimensional reduction of the RR 2-form,
$C_{2}$.
The associated Jacobi identities are (\ref{ww}) and the T-dual of
(\ref{qf}), $\omega \mathbf{F}_3=0$,
the latter ensuring the cancellation of the D5-brane flux-induced
RR charge which arises from the Chern-Simons coupling
\begin{equation}
\int_{M_4 \times \T^6} C_6 \wedge  \omega \bF_3\ ,
\end{equation}
with
 \beq (\omega X)_{pmn_1 \cdots n_{p-1}} = \omega^a_{[pm}
X_{n_1 \cdots n_{p-1} ]}{}_a   \  . \label{omX2} \eeq

\subsection{Inclusion of $H_3$ and $P$-fluxes}
\label{sec200}

In deriving the algebra (\ref{fqalgebra0}) we have applied all the
generators of $SL(2,\mathbb{Z})^7$, except for the ones of
S-duality. The latter present some subtleties which we would like
to address more carefully in this section. The aim is to extend
(\ref{fqalgebra0}) to account also for ${H_3}$ and $P$-fluxes.
These appear as $S$-duals of $F_3$ and $Q$-fluxes, respectively.
Namely,
\begin{equation}
{\cal S}_{0,2}\, F_{abc}= {H}_{abc}\ , \qquad {\cal S}_{0,2} \,
{Q}^{ab}_p = P^{ab}_p\ .
\end{equation}
Since S-duality manifests in four dimensions as electric-magnetic
duality, we find convenient to introduce new magnetic generators,
${\bar Z}_p$ and ${\bar X}^p$,
\begin{equation}
{\cal S}_{0,2}\,Z_p=  {\bar Z}_p\ , \qquad {\cal S}_{0,2}\,X^p =
 { \bar X }^p\ .
\end{equation}
In this ``democratic'' formulation of the four dimensional gauge
algebra, we expect to be able to account also for the S-duality.
Indeed, now it is straightforward to obtain the S-dual of
(\ref{fqalgebra0}),
\begin{eqnarray}
{[{\bar X}^a, {\bar X}^b]} & = &{P}^{ab}_p {\bar X}^p- {\tilde H}^{abp}{\bar Z}_p\ , \label{fwRPalgebra0}\\
{[{\bar X}^a, {\bar Z}_b]} & = &  - {P}^{ap}_b { \bar Z}_p\ ,  \nonumber\\
{[{\bar Z}_a, {\bar Z}_b]} & = &0\ , \nonumber
\end{eqnarray}
where $\tilde H_3$ is defined in a similar way than $\tilde F_3$
in eq.(\ref{tildeF}). The resulting Jacobi identities are the
S-dual expressions of (\ref{qf}) and (\ref{qq}) (see below).

The computation of commutators involving $X^a$ or $Z_a$ with their
magnetic counterparts, ${\bar X}^a $ and ${\bar Z}_a$, is more
subtle. Let us start assuming, for the sake of simplicity, that
$F_3=H_3=0$. By matching spinor indices we require the following
dependence on the Q and P-fluxes,
\begin{eqnarray}
[X^a, {\bar X}^b] & = & A\, {Q}^{ab}_p {\bar
X}^p+B\,{P}^{a \, b}_{ p }X^p \label{ztx}
\end{eqnarray}
Antisymmetry of the commutator and eq.(\ref{qq}), then gives
$A=B=1$. However, if we stick to generators in the adjoint
representation,
\begin{equation}
(X^a)^b_p={Q}^{ab}_p\ , \qquad ({\bar X}^a)_p^b= {P}^{ab}_p\ ,
%\label{}
\end{equation}
the first term in r.h.s. of (\ref{ztx}) is not independent of the
second one and the commutator can be expressed in the simpler
form,
 \begin{eqnarray}
[X^a, {\bar X}^b] & = & {Q}^{ab}_p {\bar X}^p
\end{eqnarray}
where antisymmetry is ensured by the extra condition,
 \begin{eqnarray}
 {Q}^{ab}_p {P}^{pc}_m-{P}^{ab}_p {Q}^{pc}_m=0\ .
\label{antisim}
\end{eqnarray}

Performing $SL(2,\Z)_k$  ($k=1,\dots 6$)  transformations as
above, $F_3$ and $H_3$ are reintroduced, so that the fully
$SL(2,\mathbb{Z})^7$ invariant gauge algebra associated to
Kaluza-Klein vectors is obtained\footnote{Notice also the relations between electric and magnetic generators, $Q^{ap}_b\bar Z_p-P^{ap}_b Z_p=0$ and $Q^{ab}_p\bar X^p-\tilde F^{abp}\bar Z_p-P^{ab}_p X^p+\tilde H^{abp} Z_p = 0$, ensuring the invariance of the algebra under S-duality transformations. These relations can be proven by making use of the antisymmetry conditions and the fact that the generators transform in the adjoint representation of the algebra \cite{Schon:2006kz}.}. When {\it primed}
fluxes are set to zero (see footnote \ref{pri}), it reads
\begin{eqnarray}
[X^a, X^b] & = & -\ {\tilde F}^{abp} Z_{p}+{Q}^{ab}_p X^p\ , \label{fwRPalgebra1}\\
{[X^a, Z_b]} & = &  -{Q}^{ap}_b Z_p\ , \nonumber\\
{[X^a, {\bar X}^b]} & = & {Q}^{ab}_p {\bar X}^p-\ {\tilde F}^{abp}{\bar Z}_p\ , \nonumber\\
{[{\bar X}^a, Z_b ]} & = &  {[{X}^a, {\bar Z}_b ]} = -{Q}^{ap}_b {\bar Z}_p\ , \nonumber\\
{[{\bar X}^a, {\bar X}^b]} & = &{P}^{ab}_p {\bar X}^p- {\tilde H}^{abp}{\bar Z}_p\ , \nonumber\\
{[{\bar X}^a, {\bar Z}_b]} & = &  - {P}^{ap}_b { \bar Z}_p\ ,\nonumber\\
{[{\bar Z}_a, {\bar Z}_b]} & = &{[Z_a, Z_b]}={[Z_a, {\bar
Z}_b]}=0\ . \nonumber
\end{eqnarray}
As before, the constraints arising from the Jacobi identities of
the algebra,
\begin{eqnarray}
{Q}^{[ab}_p{Q}^{c]p}_l& = &0\ , \label{fwRPJI1}\\
{P}^{[ab}_p{P}^{c]p}_l & = &0\ , \label{fwRPJI2}\\
 {Q}^{[ab}_p{P}^{c]p}_l={P}^{[ab}_p{Q}^{c]p}_l & = &0\
 ,\label{fwRPJI3}\\
 {Q}^{[ab}_p{\tilde F}^{c]lp}+{\tilde F}^{p[ab} {Q}^{c]l}_p& = & 0\quad \Leftrightarrow \quad QF_3=0\
 ,\label{fwRPJI4}\\
{\tilde H}^{p[ab}{P}^{c]l}_p+{P}^{[ab}_p {\tilde H}^{c]lp}& = & 0\quad \Leftrightarrow \quad PH_3=0\ , \label{fwRPJI5}\\
\begin{minipage}{5cm}\begin{eqnarray*}&&{Q}^{l[a}_p {\tilde H}^{bc]p}-{P}^{[ab}_p{\tilde F}^{c]lp}=\\
&&\qquad{P}^{l[a}_p {\tilde F}^{bc]p}-{Q}^{[ab}_p{\tilde H}^{c]lp}=
0\end{eqnarray*}\end{minipage}&& \Leftrightarrow \ \
\begin{cases}
PF_3+QH_3=0\\
Q^{l[a}_p\tilde{H}^{bc]p}+Q^{[ab}_p\tilde{H}^{c]lp}-P^{l[a}_p\tilde{F}^{bc]p}-P^{[ab}_p\tilde{F}^{c]lp}=0
\end{cases} \label{fwRPJI6}
\end{eqnarray}
must be supplemented with eq.(\ref{antisim}) and,
\begin{eqnarray}
-{Q}^{ab}_p {\tilde H}^{clp}+{P}^{ab}_p\,{\tilde F}^{clp}+ {\tilde
H}^{pab}\, {Q}^{cl}_p -{\tilde F}^{pab}{P}^{cl}_p & = &0\ ,
\end{eqnarray}
ensuring the antisymmetry of mixed generators. We have recovered
in this way, both, the $(p,q)$ 7-brane charge tadpole cancellation
conditions, given in eqs.(\ref{7can})-(\ref{ns7can}), and the
Bianchi identities, eqs.(\ref{fwRPJI1})-(\ref{fwRPJI3}), which Q
and P-fluxes must satisfy for consistency.

Equations (\ref{fwRPJI1}), (\ref{fwRPJI2}) and a constraint analogous to the last condition
in (\ref{fwRPJI6}), were already conjectured in \cite{acfi} by
symmetry arguments\footnote{The last condition in (\ref{fwRPJI6}) that we have derived here is actually slightly weaker than the singlet condition conjectured in \cite{acfi}.}. In addition, a constraint
${Q}^{[ab}_p{P}^{c]p}_l+{P}^{[ab}_p{Q}^{c]p}_l=0$ was also
conjectured. This has been recently interpreted as a cohomology
condition in the deformations of (\ref{xxq}) by its second
cohomology class \cite{guarino}. Here, however, we observe a
stronger requirement, given by eq.(\ref{fwRPJI3}). Not every
consistent deformation of the algebra (\ref{xxq}) appears as a
valid deformation. The ten dimensional origin of this constraint,
having a RR structure, is unknown to us. However, it is worth
noticing that (using identifications to be discussed below)
couplings proportional to ${Q}^{ab}_p {P}^{pc}_m$ are found in
topological terms of $\mathcal{N}=4$ supergravity actions (see for
instance \cite{Schon:2006kz}) coupled to gauge vector fields,
$\epsilon ^{\mu \nu \rho \lambda} A_{\mu a}^+A_{\nu b }^-A_{\rho c
}^+A_{ \lambda ^m }^-$.

%%%%%%%%%%%%%%%%%%%%%%%%%%%%%%%%%%%%%%%%%%%%%%%%%%%%%%%%%%%%%%%%%%%%%%

\section{Brane gaugings and Freed-Witten anomalies}
\label{sec4}

The algebra derived in the previous section is incomplete in
several aspects which concern the twisted states of the theory.
First, it does not include gauge symmetries arising at loci where
the elliptic fiber degenerates. That is, the gauge degrees of
freedom in the worldvolume of the 7-branes. Second, it does not
contain information on sectors of the theory which are not aligned
with 7-brane charges, such as the D3-brane tadpole cancellation
requirements.

In this section we address the first of these points, whereas the
second one will be partially addressed in the next section.

To illustrate how chiral configurations of branes can be described
in the gauging formalism, we find convenient to start considering
dual type I compactifications on twisted tori, as described at the
end of section \ref{sec21}, with magnetized D9-branes
\cite{magnetized}. The magnetization simply amounts to introducing
some worldvolume fluxes, which can be described as gaugings.
Namely, due to the Chern-Simons couplings on the worldvolume of
the branes, certain RR axionic scalars shift under D-brane $U(1)$
gauge symmetries (due to the presence of the magnetic fields).
This is in fact a gauging.

A key point is that configurations of magnetized D9-branes have
non-trivial Freed-Witten (FW) constraints in the presence of bulk
fluxes. Some D9-brane configurations are not allowed in the
presence of worldvolume fluxes, showing a sort of mutual
consistency between open and closed string backgrounds. In
particular, in \cite{cfi} (see also \cite{vz,oscar,vz2}) it was
suggested that FW constraints could be understood as the
requirements needed to ensure invariance of the effective
superpotential under shifts of the  $U(1)$ gauged axionic scalars.

Here we propose to study this interrelation between open and
closed string fluxes from the point of view of the complete
algebra describing both, bulk and worldvolume gaugings.  The idea
is that FW anomaly cancellation requirements arise as a Jacobi
identity for the algebra involving both, the closed string
generators and the open string ones associated to $U(1)$ gauge
fields on the D9-brane worldvolume.

The final aim will be to extend the algebra (\ref{fwRPalgebra1})
to account for the gauge generators associated to $(p,q)$
7-branes. As it will be clear below, once these are included, the
tadpole cancellation conditions, involving localized 7-brane
sources with non-zero total charge, and the Freed-Witten anomaly
cancellation conditions, arise from the Jacobi identities of the
algebra.

\subsection {A simple example}

Let us consider  a configuration of D6-branes in the presence of
NSNS $H_3$ flux. The D6-branes wrap 3-cycles of the internal
factorized 6-torus, with wrapping numbers
\begin{equation*}
(n_1,m_4)\times (n_2,m_5)\times (n_3,m_6)\ .
\end{equation*}
For the sake of simplicity, we turn on only some particular
component of $H_3$, say $\ov{H}_{423}$. The FW constraint in this
case simply leads to impose \cite{cfi}
 \beqa m_4n_2n_3
\ov{H}_{423}=0 \quad \quad {i=1,2,3} \label{fw}\eeqa

Under three T-dualities along the directions $x^4$, $x^5$ and
$x^6$, $\ov{H}_{423}$ maps to a metric flux, $-\omega^4_{23}$ (see
Table \ref{nongeoiib}), specifying that the $S^1$ parameterized by
$x^4$ is non-trivially fibered over the 2-torus $T^2_{[x^2x^3]}$,
with first Chern class given by the integer $\ov{H}_{423}$. Hence,
moving around the rectangle spanned by $x^2$ and $x^3$ implies
picking up a shift in $x^4$. This intuition is formalized in the
algebra relation between the generators of shifts $Z_i$ \beqa [
Z_2,Z_3 ]=\omega _{23}^4 Z_4 \eeqa

The Freed-Witten  constraint (\ref{fw}) has a clear interpretation
in this picture (see \cite{marchtorsion}). The quantity $n_2 n_3
m_4$ counts the charge of D5-branes wrapping $T^2_{[x^2x^3]}$. But
the space parameterized by $x^2$ and $x^3$ is no longer a valid
2-cycle: because of the above mentioned shift, it has a $S^1$
boundary parameterized by $x^4$.

These nice geometric interpretations, however, are not always
available, so we would like to move on towards an algebraic
characterization of such constraints, based on consistency
conditions of the gauge algebras.

Let us be a bit specific in our example. By defining
$B_2=\int_{T^4_{[x^2x^3x^5x^6]}} C_6$ and its dual,
$a=\int_{T^2_{[x^1x^4]}} C_2$, we can write the D9 Chern-Simons
coupling
 \beqa \int_{D9} C_6\wedge F \wedge F = m_4n_2n_3
\int_{4d} B_2 \wedge F\ , \eeqa or, \beqa m_4n_2n_3 \int_{4d}
A^\mu
\partial_\mu a\ , \eeqa
where $n_2$ ($n_3$) is the number of times the cycle spanned by
$x^2$ ($x^3$) is wrapped by the brane, and $\int_{T_{[x^1x^4]}^2}
F=\frac{m_{4}}{n_1}$ is the magnetic flux in the first 2-torus
\cite{magnetized}. Thus, as mentioned,  the isometry corresponding
to a shift in $a$ is gauged by the $U(1)$ gauge transformation of
the D9-brane. It is therefore natural to introduce a generator,
$X$, in the four dimensional theory associated to these $U(1)$
gauge transformations, and to write down the commutators of $X$
with the generators arising from the closed string modes.

The $U(1)$ gauge bundle is non-trivially fibered over the first
torus, with first Chern class given by the integer $m_4 n_2 n_3$.
In analogy with the geometric flux above, there is a gauge
transformation when going around the rectangle in $x^1$ and $x^4$.
That is, we should have \beqa [Z_1,Z_4]=m_4n_2n_3 X\ , \eeqa with
all other commutators vanishing.

It is easy to see, then, that the corresponding FW constraint
\beqa
 \omega_{23}^4 m_4n_2n_3 = 0\ ,
\eeqa results from the Jacobi identities of the algebra.

\subsection{Freed-Witten conditions and gaugings for magnetized D9-branes}

Let us now generalize the type I example of the previous section.
For that, we consider a stack of D9 branes, wrapping $n_i$ times
the 2-cycle $[\omega_i]$, $i=1,2,3$, and magnetic flux given by
\begin{equation}
{n_i}\, {\cal F}_{i,i+3}\equiv {n_i} \, \int_{[\omega_i]}
F=m_{i+3}\ . \label{magneticf}
\end{equation}
In type IIB with O9-planes and D9 magnetized branes, there are
Chern-Simons couplings of the gauge field strength to the RR
forms,
\begin{eqnarray}
\int_{D9} C_6\wedge  F \wedge  F = c_i \int_{4d} B^i_2\wedge  F \
, \qquad \int_{D9} C_2\wedge F \wedge F\wedge   F \wedge  F= c_0
\int_{4d} C_2\wedge F\ ,\label{coup}
\end{eqnarray}
where $c_i$ is the corresponding first Chern class,
\begin{eqnarray}
&& c_0=-n {\cal F}_{14} {\cal F}_{25}{\cal F}_{36}=- m_4m_5m_6\ , \\
&& c_1=n {\cal F}_{14}= m_4 n_2 n_3\ , \\
&& c_2=n {\cal F}_{25}=m_5 n_1 n_3 \ ,\\
&& c_3=n {\cal F}_{36}=m_6 n_1 n_2\ . \label{gaugecoupling}
\end{eqnarray}
In these expressions, $n=n_1n_2n_3$, and we have defined the four
dimensional 2-forms
\begin{equation}
B^i_2=\int_{[\tilde \omega_i]} C_6\ .  \label{cs}
\end{equation}
These are indeed all the couplings of $F$ to 2-forms allowed by
the orientifold projection. Introducing also the four dimensional
Hodge duals,
\begin{eqnarray}
&& da^i= *_4 dB^i_2 \ ,  \quad \quad da^0= -*_4 dC_2\ ,
\end{eqnarray}
where the fields $a^I$, with $I=0\ldots 3$, correspond
respectively to the RR axions, Im $S$, Im $T_i$, the couplings
(\ref{coup}) can be rewritten as,
\begin{eqnarray}
&& c_i \int \partial_{\mu}a^i A^{\mu} \ , \quad \quad c_0 \int
\partial_{\mu}a^0 A^{\mu}
\end{eqnarray}
Hence, a shift $a^i\rightarrow  a^i+c_i\chi$ in the scalar fields
is gauged by the gauge transformation of the diagonal $U(1)$ on
the D9-branes. Generalizing the previous example and introducing a
gauge generator $X$ for this $U(1)$, we write down the algebra,
\begin{eqnarray}
{[Z_a, Z_b]} & = &  {\omega}_{ab}^p Z_p  +n {\cal F}_{ab}X + \dots
\end{eqnarray}
where the dots refer to other possible bulk fluxes. Actually, we
can do it better and consider multiple $U(1)$ subgroups within the
D9-brane gauge group, each one with a generator $X^I$. Proceeding
as in section \ref{sec20}, we obtain the following gauge algebra,
\begin{eqnarray}
{[Z_a, Z_b]} & = &  {\omega}_{ab}^p Z_p  +n{\cal F}_{ab}^IX^I -\bF_{abp} Y^ p \ , \label{hetdual1}\\
{[Z_a, Y^p]}& = &  -{\omega}_{aq}^p Y^ q\ , \nonumber \\ {[Z_a,
X^I]}& = & {\cal F}_{aq}^I Y^ q \ , \nonumber
\end{eqnarray}
with Jacobi identities
\begin{eqnarray}
{\omega}_{[ab}^p {\omega}_{c]p}^l& = &0\ ,\\
  {\omega}_{[ab}^p n {\cal F}^I_{c]p} & = &0\ ,\label{fww}\\
  {\bF}_{p[ab}   {\omega}_ {c]l}^p+  {\omega}_{[ab}^p {\bF}_{c]lp}&= &- n \sum_I{\cal F}^I_{[ab}{\cal
  F}^I_{c]l}\ .\label{tad}
\end{eqnarray}
This algebra could has been obtained, alternatively, by type
I/heterotic duality, from the heterotic algebra found in
\cite{Kaloper:1999yr}, after adjusting some normalization factors,
and identifying the worldvolume magnetic fluxes with abelian
Yang-Mills fluxes in the heterotic side.

Let us comment on the new equations, (\ref{fww}) and (\ref{tad}).
The first one, is the standard Freed-Witten anomaly cancellation
requirement for magnetized D9-branes. Indeed, it is possible to
check that it ensures the invariance of the effective
superpotential
 \beq W_{O9} = \int_{\T^6}  \Omega \wedge
(F_3+\omega J_c)   \ , \label{wbfhq} \eeq under the shifts
\begin{equation}
J_c  \rightarrow  J_c + n\mathcal{F}  \chi\ ,
\end{equation}
with $J_c=C_2+iJ$.

Equation (\ref{tad}), on the other hand, was shown in section
\ref{sec20} to correspond to the tadpole cancellation requirement
for the D5-brane charge. Now, however, the condition appears
sourced by a new term, corresponding to the solitonic charge of
D5-brane induced in the worldvolume of the magnetized D9-branes
\begin{equation}
\int_{M_4 \times \T^6} C_6  \wedge \omega \bF_3  -\int_{D9}  C_6
\wedge F\wedge F\ .
\end{equation}
In this way, as advanced, the inclusion of new generators in the
closed string algebra, corresponding to diagonal $U(1)$ gauge
symmetries coming from the open string sector, leads to the
correct Freed-Witten anomaly and tadpole cancellation conditions.

In what follows, we apply this procedure to extend the algebra
(\ref{fwRPalgebra1}) in order to account also for possible $(p,q)$
7-branes.

\subsection{Gaugings and Freed-Witten conditions for F-theory 7-branes}

Let us come back to the main setup considered in this work, i.e.
globally non-geometric compactifications of F-theory, and consider
the presence of magnetized $(p,q)$ 7-branes, containing an ADE
gauge theory in their worldvolume. As we have seen, there are
generators associated to the $U(1)$ subgroups of these gauge
theories which play an important role in the algebra. We will try
to implement these by making use of T and S-dualities.

Type IIB magnetized D$7_k$-branes can be seen as the T-dual of
magnetized D9-branes, with magnetization numbers,
\begin{align*}
D7_1:\qquad (1,0)\times (n_2,m_5)\times (n_3,m_6)\ , \\
D7_2:\qquad (n_1,m_4)\times (1,0)\times (n_3,m_6)\ , \\
D7_3:\qquad (n_1,m_4)\times (n_2,m_5)\times (1,0)\ .
\end{align*}
where six T-dualities are applied along the internal 6-torus. The
integers $n_i$ and $m_i$ now correspond to the magnetization of
the $D7_k$-branes, defined accordingly to,
\begin{equation}
m_{i+3}\int_{[\omega_i]}F = n_i\ ,
\end{equation}
with $[\omega_i]\subset [\tilde \omega_k]$.

From the algebra (\ref{hetdual1}) it is then simple to obtain the
dual type IIB algebra,
\begin{eqnarray}
[X^a, X^b] & = & -\ {\tilde F}^{abp} Z_{p}+{Q}^{ab}_p X^p+(c^k_I)^{ab}X^I_k\ , \label{fqopen}\\
{[X^a, Z_b]} & = &  -{Q}^{ap}_b Z_p\ , \nonumber\\
{[Z_a, Z_b]} & = &[Z_a, X^I_k]=[X^I_k, X^J_j]=0\ , \nonumber\\
{[X^a, X^I_k]} & = &(\mathcal{F}^I_k)^{aq}Z_q\ , \nonumber
\end{eqnarray}
corresponding to the extension of the algebra (\ref{fqalgebra0}).
The index $k=1,2,3$ labels the three 2-torus transverse to the
D$7_k$-branes, whereas $I$ runs over the different stacks of
branes. The first Chern class associated to each of the stacks is
now encoded in the quantities
\begin{equation}
c^1_I=n^I_2n^I_3\mathcal{F}^I_1\ , \qquad
c^2_I=n^I_1n^I_3\mathcal{F}^I_2\ , \qquad
c^3_I=n^I_1n^I_2\mathcal{F}^I_3\ ,\label{cher7}
\end{equation}
where
\begin{equation}
(\mathcal{F}^I_k)_{a,a+3}=-(\mathcal{F}^I_k)_{a+3,a}=\frac{m^I_{a+3}}{n^I_a}
\qquad \textrm{for }k\neq a\ ,\label{cher7b}
\end{equation}
and zero otherwise.

We would like to extend this algebra to account also for $H_3$,
$P$-fluxes and general $(p,q)$ 7-branes. For that, we follow the
same procedure than in section \ref{sec200}, and introduce
magnetic dual $U(1)$ generators, $\bar X^I$, associated to the
gauge fields in the worldvolume of the 7-branes. The existence of
two worldvolume vectors is consistent with the fact that both, F
and D-strings, can end on the 7-brane \cite{7inv}. The diagonal
$U(1)$ gauge symmetry in the worldvolume of a stack of $(p,q)$
$7_k$-branes is generated by the linear combination
\begin{equation}
p^I_k X^I_k+q^I_k \bar X^I_k\ .
\end{equation}

Starting with the algebra (\ref{fqopen}) and acting with the
S-duality generators, as in section \ref{sec200}, we obtain the
extension of the full algebra (\ref{fwRPalgebra1}), which accounts
also for the gauge symmetries associated with the $(p,q)$
$7_k$-branes present in the compactification,
\begin{eqnarray}
[X^a, X^b] & = & -\ {\tilde F}^{abp} Z_{p}+{Q}^{ab}_p X^p+\sqrt{2}\ p^I_k (c^k_I)^{ab}X^I_k\ , \label{algfinal}\\
{[X^a, Z_b]} & = &  -{Q}^{ap}_b Z_p\ , \nonumber\\
{[X^a, X^I_k]} & = &\sqrt{2}\ p^I_k (\mathcal{F}^I_k)^{aq}Z_q\ , \nonumber\\
{[X^a, {\bar X}^b]} & = & {Q}^{ab}_p {\bar X}^p-\ {\tilde F}^{abp}{\bar Z}_p+\sqrt{2}\ p^I_k (c^k_I)^{ab}\bar X^I_k\ , \nonumber\\
{[{\bar X}^a, Z_b ]} & = &  {[{X}^a, {\bar Z}_b ]} = -{Q}^{ap}_b {\bar Z}_p\ , \nonumber\\
{[X^a, \bar X^I_k]} & = &{[{\bar X}^a, X^I_k]}=\sqrt{2}\ p^I_k (\mathcal{F}^I_k)^{aq}\bar Z_q\ , \nonumber\\
{[{\bar X}^a, {\bar X}^b]} & = &{P}^{ab}_p {\bar X}^p- {\tilde H}^{abp}{\bar Z}_p+\sqrt{2}\ q^I_k (c^k_I)^{ab}\bar X^I_k\ , \nonumber\\
{[{\bar X}^a, {\bar Z}_b]} & = &  - {P}^{ap}_b { \bar Z}_p\ ,\nonumber\\
{[{\bar X}^a, \bar X^I_k]} & = &\sqrt{2}\ q^I_k (\mathcal{F}^I_k)^{aq}\bar
Z_q\ , \nonumber
\end{eqnarray}
with all the other commutators vanishing. In addition, the
following conditions are also required to ensure antisymmetry of
the commutators,
\begin{align}
&{Q}^{ab}_p {P}^{pc}_m-{P}^{ab}_p {Q}^{pc}_m=0\ , \\
&-{Q}^{ab}_p {\tilde H}^{clp}+{P}^{ab}_p\,{\tilde F}^{clp}+ {\tilde
H}^{pab}\, {Q}^{cl}_p -{\tilde F}^{pab}{P}^{cl}_p= 0 \ , \\
&(\mathcal{F}^I_k)^{ap}(q^I_k Q^{cd}_p-p^I_k P^{cd}_p)= 0\ .
\end{align}
The Jacobi identities of the algebra can be then summarized in the
following sets of constraints,
\begin{itemize}
\item[-] \emph{Bianchi identities:}
\begin{align}
&{Q}^{[ab}_p{Q}^{c]p}_l = 0\ , \label{bian1}\\
&{P}^{[ab}_p{P}^{c]p}_l  = 0\ , \\
&{Q}^{[ab}_p{P}^{c]p}_l={P}^{[ab}_p{Q}^{c]p}_l  = 0\
 ,\\
&Q^{l[a}_p\tilde{H}^{bc]p}+Q^{[ab}_p\tilde{H}^{c]lp}-P^{l[a}_p\tilde{F}^{bc]p}-P^{[ab}_p\tilde{F}^{c]lp}=0\ ,
 \label{bian2}
 \end{align}
\item[-] \emph{7-brane tadpoles:}
\begin{align}
&(QF_3)_k=-\sum_I (p_k^I)^2 d_k^I\
 ,\label{tad1}\\
&(PH_3)_k=-\sum_I (q_k^I)^2 d_k^I\ , \\
&(PF_3+QH_3)_k=-2\sum_I p_k^Iq_k^I d_k^I \ ,\label{tad2}
\end{align}
\item[-] \emph{Freed-Witten anomalies:}
\begin{align}
Q^{[ab}_p(c^k_I)^{c]p}=0\ , \\
P^{[ab}_p(c^k_I)^{c]p}=0\ ,
\end{align}
\end{itemize}
where we have made use of the wrapping numbers of the $I$-th stack
of $7_k$-branes, $d_k^I$, defined as
\begin{equation}
d_1^I=m_5^Im_6^I\ , \qquad d_2^I=m_4^Im_6^I\ ,\qquad
d_3^I=m_4^Im_5^I\ .
\end{equation}
Notice that, for non-vanishing $H_3$ and $P$-fluxes, the
superpotential (\ref{pqsuper}) also develops a linear dependence
in the axion-dilaton. Invariance under the shifts
\begin{equation*}
\textrm{Im }S\ \to \ \textrm{Im }S + c_0\chi\ ,
\end{equation*}
would lead to Freed-Witten anomaly cancellation requirements for
magnetized D9-branes with $c_0\neq 0$. Here, however, we only
consider the presence of 3 and 7-branes (see footnote \ref{d9}).

We have obtained, thus, the full gauge algebra associated to
Kaluza-Klein and 7-brane vectors in the class of F-theory
compactifications described in section \ref{sec1}. The Jacobi
identities of the algebra encode all the information relative to
the Bianchi identities, tadpole cancellation conditions for
$(p,q)$ 7-branes and Freed-Witten anomaly cancellation
requirements, corresponding to the electric and magnetic parts of
the worldvolume fields. Hence, whereas for the class of fluxes we
are considering here, the consistency conditions which the
background must satisfy are elusive in a ten dimensional
description in terms of local supergravity solutions plus stringy
monodromies, the four dimensional gauge algebra captures much of
the information of the higher dimensional model. The underlying
philosophy is that every consistent (holomorphic) gauging of the
four dimensional effective supergravity is capturing the dynamics
of some light modes of the higher dimensional superstring
theory\footnote{As pointed out in \cite{scan}, it may well happen,
however, that the higher dimensional solution does not correspond
to a compact manifold.}, even  if the latter is in the
strongly coupled and/or $\alpha'\gg 1$ regime. Of course, in the
latter case, corrections to the effective supergravity description
are expected to be important.

%%%%%%%%%%%%%%%%%%%%%%%%%%%%%%%%%%%%%%%%%%%%%%%%%%%%%%%%%%%%%%%%%%%%%%%%%

\section{Comparison with gauged supergravity actions}
\label{sec3}

\subsection{$\mathcal{N}=4$ gauged supergravity}

Before the inclusion of fluxes, the effective four dimensional
theory resulting from orientifold compactification on a 6-torus
preserves $\mathcal{N}=4$ supersymmetry. Such is the case for
example of F-theory compactified on $K3\times T^4$, discussed in
\cite{sen}.

The covariant formulation of $\mathcal{N}=4$ gauged supergravity
in four dimensions has been proposed in \cite{Schon:2006kz}. The
big amount of local supersymmetry constrains the only possible
deformations of the theory to be gaugings induced by minimal
couplings of vector fields to ``isometry'' generators of the
background. The structure constants of the gauge algebra are
naturally identified with flux parameters of the higher
dimensional compactified theory, as depicted in the previous
sections.

The set of possible gaugings of $\mathcal{N}=4$ supergravity, with
$n$ extra vector multiplets, is encoded in two embedding tensors
of $SL(2,\mathbb{Z})\times SO(6,6+n;\mathbb{Z})$
\cite{Schon:2006kz}
\begin{equation}
f_{\alpha MNP}\ , \qquad \xi_{\alpha M}
\end{equation}
with $\alpha=\pm 1/2$ and $M=1\ldots 12+n$. Supersymmetry and
anomaly cancellation requires the emergence of a potential for the
scalars living in the vector multiplets
\begin{multline}
V_{\mathcal{N}=4}= -\frac{1}{16}\left[f_{\alpha MNP}f_{\beta
QRS}M^{\alpha\beta}\left(\frac13
M^{MQ}M^{NR}M^{PS}+\left(\frac23\eta^{MQ}-M^{MQ}\right)\eta^{NR}\eta^{PS}\right)\right.\\
\left.-\frac49 f_{\alpha MNP}f_{\beta
QRS}\epsilon^{\alpha\beta}M^{MNPQRS}+3\xi_\alpha^M\xi_\beta^NM^{\alpha\beta}M_{MN}\right]\
,\label{scag}
\end{multline}
together with a topological term for the vector fields
\cite{Schon:2006kz}. The indices are lowered and raised with the
off-block diagonal metric,
\begin{equation}
\eta_{MN}=\eta^{MN}=\begin{pmatrix}0&\mathbb{I}_6\\
\mathbb{I}_6& 0\end{pmatrix}\ ,\label{eta}
\end{equation}
and the antisymmetric tensor
$\epsilon^{\alpha\beta}=\epsilon_{\alpha\beta}=\pm 1$. The
matrices $M_{MN}$ and $M_{MNPQRS}$ are given in terms of a
$SO(6,6)$ vielbein matrix $\mathcal{V}^P{}_Q$ as
\begin{equation}
M_{MN}=\mathcal{V}^P{}_M\mathcal{V}^P_N\ , \quad
M_{MNPQRS}=\epsilon_{mnopqr}\mathcal{V}^m{}_M\mathcal{V}^n_N\mathcal{V}^o{}_P\mathcal{V}^p_Q\mathcal{V}^q{}_R\mathcal{V}^r{}_S\
,\label{m6}
\end{equation}
where the lower-case indices run from 1 to 6.\footnote{Notice that vielbeins in eq.(\ref{m6}) are rotated with respect to those in Ref.\cite{Schon:2006kz} since we are using a different, off-diagonal, $SO(6,6)$ metric.} In the minima of the
scalar potential some of the scalars acquire a mass, reflecting
the moduli stabilization induced by the fluxes.

Let us consider first the simplest case  consisting on pure
super-Poincar\'e symmetry with no extra vector multiplets ($n=0$).
This is exactly the same situation discussed in section
\ref{sec2}, where the $U(1)$ generators of the branes were
decoupled from the Kaluza-Klein vectors. In that case, the
matrices $M^{MN}$ and $M^{\alpha\beta}$ can be chosen to be
parameterized as
\begin{equation}
M^{MN}=\begin{pmatrix}g^{mn}&-g^{mk}c_{kn}\\
c_{mk}g^{kn}&g_{mn}-c_{mk}g^{k\ell}c_{\ell n}
\end{pmatrix}\ , \quad M^{\alpha\beta}=\frac{1}{\textrm{Re
}S}\begin{pmatrix}1&\textrm{Im }S\\
\textrm{Im }S&|S|^2\end{pmatrix}\ ,
\end{equation}
where, the tensors $g_{mn}$ and $c_{mn}$ correspond respectively
to the metric tensor and the RR 4-form of a toroidal
compactification of type IIB supergravity,
\begin{equation}
ds^2=\frac12g_{mn}dx^mdx^n \ , \qquad C_4=\frac12
c_{mn}*_6(dx^m\wedge dx^n)\ .
\end{equation}
For the particular case of a factorizable torus, these are given
in eqs.(\ref{metrica}) and (\ref{c4}). The $\mathbb{Z}_2\times
\mathbb{Z}_2$ symmetry ensuring the factorization of the
background breaks the $SO(6,6)\times SL(2,\mathbb{Z})$ global
symmetry group to a $[SL(2,\mathbb{Z})]^7\subset SO(7,7)$
subgroup, as already described. The action on the embedding tensor
is such that $\xi_{\alpha M}$ is projected out completely, whereas
only the components
\begin{equation}
(M \ \textrm{mod} \ 3) \ \neq \ (N \ \textrm{mod} \ 3) \ \neq \ (P
\ \textrm{mod} \ 3)\label{trunc}
\end{equation}
of $f_{\alpha MNP}$ survive to the projection. The gauge group
generators are then easily identified in terms of the embedding
tensor, and the full gauge algebra is given by
\cite{Schon:2006kz}
\begin{eqnarray}
[Z_{\alpha A}, Z_{\beta B}] & = & \delta_{\beta}^{\rho} f_{\alpha
A B}^P Z_{\rho P}\ ,\label{gaugedalg}
\end{eqnarray}
leading to general Jacobi identities,
\begin{eqnarray}
f_{\alpha [A B}^P f_{\beta B]P}^L=0\label{jaco1}\\
f_{+ A B}^P f_{- BP}^L-f_{- A B}^P f_{+ BP}^L=0\label{jaco2}
\end{eqnarray}
The algebra (\ref{fwRPalgebra1}), which we have constructed by
using modular transformations, exactly matches (\ref{gaugedalg}),
with the extra flux parameters corresponding to primed fluxes,
which we have set to zero (see footnote \ref{pri}). This
constitutes a non trivial crosscheck that non-geometric flux
compactifications, like the ones described in section \ref{sec1},
lead to consistent gauged supergravities in four dimensions. More
precisely, the following dictionary between the flux spinor
$\mathbb{G}$ and the projected in components of the embedding
tensor can be established
\begin{equation}
\mathbb{G}_{(s_0,s_1,s_2,s_3,s_4,s_5,s_6)}=\textrm{Sign}(s_0M_1M_2M_3)\
f_{s_0, |M_1|, |M_2|, |M_3|}\label{dic}
\end{equation}
where $s_i=\pm 1/2$ and
\begin{multline}
M_k = \frac14[(k+9)(2s_{k}-1)(2s_{k+3}-1)\ - \
(k+6)(2s_{k}-1)(2s_{k+3}+1)\ +
\\ + \ (k+3)(2s_{k}+1)(2s_{k+3}+1)\ +\
k(2s_{k}+1)(2s_{k+3}-1)]\ .
\end{multline}
Equivalently, in terms of $[SO(2)]^3$ covariant tensors,
\begin{align}
f_{+ a}{}^{bc} &= - \frac12 Q^{bc}_a\ , \label{flux1} &
f_{+}{}^{abc} &=  \frac12 \tilde F^{abc}\ , \\
f_{-a}{}^{bc} &= - \frac12 P^{bc}_a\ , &
f_-{}^{abc} &=  \frac12 \tilde H^{abc}\ , \label{flux2}
\end{align}
and similarly for the primed fluxes. Here, $a,b,c=1\ldots 6$, and
indices are raised with the $SO(6,6)$ metric $\eta_{MN}$, given in
eq.(\ref{eta}). Notice that electric (magnetic) gaugings do not
necessarily correspond to NSNS (RR) fluxes, contrary to what it is
sometimes stated in the supergravity literature.

We can now easily extend this dictionary to the case on which
$(p,q)$ 7-branes are also present. Notice that $\mathcal{N}=4$
supersymmetry allows, at most, for a single class of 7-branes
wrapping one of the three 4-cycles of the factorized 6-torus.
There are electric and magnetic generators, $X^I$ and
$\overline{X}^I$, $I=0\ldots n$, corresponding to diagonal $U(1)$
symmetries in the worldvolume of the the 7-branes. The uppercase
indices of the embedding tensor, $f_{\alpha MNP}$, now run from 1
to $12+n$. And, in particular, looking at the structure of
(\ref{algfinal}), we see that we can accommodate the 7-brane
gaugings in the following components of the embedding tensor,
\begin{equation}
\frac12(c_I)^{ab}=-\frac{1}{p^I}
f_{+(12+I)}{}^{ab}=-\frac{1}{q^I}
f_{-(12+I)}{}^{ab}\ ,
\end{equation}
where $(c_I)^{ab}$ is the gauging parameter defined in
(\ref{cher7}) and (\ref{cher7b}), for the relevant 7-brane.

\subsection{$\mathcal{N}=1$ structure and D3-brane sector}

We have just observed that the algebra (\ref{fwRPalgebra1}) fits
perfectly within a truncation of a $\mathcal{N}=4$ gauged
supergravity algebra. This is because in deriving
(\ref{fwRPalgebra1}) we have actually made use of the symmetries
and dualities of $\mathcal{N}=4$ supergravity. We therefore, do
not expect the algebra (\ref{fwRPalgebra1}) to contain intrinsic
information of the $\mathcal{N}=1$ supersymmetry. A first hint of
this is that the tadpole cancellation condition for the D3-brane
charge, given in eq.(\ref{d3tadpole}), does not arise from the
quadratic constraints, eqs.(\ref{jaco1})-(\ref{jaco2}), of the
algebra. Indeed, D3-branes are special in that they couple
electrically to the RR 4-form, $C_4$, which only fills the
non-compact directions. In this sense, the cancellation of the
D3-brane charge is not a topological condition which can be
derived from the algebra (\ref{algfinal}).

A detailed analysis of the scalar potential (\ref{scag}) moreover shows that the truncation (\ref{trunc}) together
with the Jacobi identities of the $\mathcal{N}=4$ algebra guarantee a
$\mathcal{N}=1$ structure for (\ref{scag}),\footnote{We thank the authors of \cite{adolfo} for noticing an error in previous versions which was leading to an apparent disagreement between both scalar potentials.}
\begin{equation}
V_{\mathcal{N}=1}=
e^K\left(\sum_{\mathbb{T}_i}(\mathbb{T}_i+\bar{\mathbb{T}}_i)^2|D_{\mathbb{T}_i}W|^2-3|W|^2\right)\
,\label{scas}
\end{equation}
with $K$ the K\"ahler potential of the factorized torus, given in
eq.(\ref{kahler}) and $W$ the superpotential in eq.(\ref{pqsuper}).

%%%%%%%%%%%%%%%%%%%%%%%%%%%%%%%%%%%%%%%%%%%%%%%%%%%%%%%%%%%%%%%%%%%%%%%%%%%

\section{Summary and  Outlook}
\label{sec5}

The main aim of the  this work was to advance in the understanding
of string compactifications when fluxes, geometric and
non-geometric, are switched on. Different dualities suggest the
presence of a huge collection of background fluxes. For instance,
the setup considered in our work  (where an initial orientifold
compactification on factorized tori is enforced) allows for as
many as  $2^7$ of such fluxes,   encoded in $SL(2,\Z)^7$ spinorial
representations. Flux parameters manifest as gaugings in four
dimensions. The gauged four dimensional supergravity theory
contains information about the fully stringy aspects of the
original theory and, generically, it is not just a
compactification of ten dimensional supergravity.  As a
consequence, identifying the  ten dimensional origin of the
gaugings is far from being straightforward. One goal of our study
was to show that half of these fluxes can be associated  to
global, non-geometric, compactifications of F-theory. The
compactification manifold appears to be an  U-fold
\cite{ufolds,ufolds1} where local patches are glued by performing
T and S-duality transformations. We illustrated such ideas through
an explicit example, which, by using  S and  T dualities, can be
mapped into a standard Type IIB compactification.

Comprehension of the algebraic gauge structure of the effective
theories arising from superstring theory compactifications is of
great importance for understanding both, the four dimensional
physics and the vacuum structure of the higher dimensional theory.
By means of a systematic use of duality transformations, we were
able to derive the full ${\cal N}=4$ algebra satisfied by gauge
generators related to Kaluza-Klein vector fields. Jacobi
identities of such algebra provide a consistent way of finding
constraints satisfied by fluxes.

Moreover, we have shown that the 7-brane sector (or more
generically, the open string sector or twisted sector) can also be
settled in a similar framework, where branes are described in the
gauging formalism. We found that the Jacobi identities of the
algebra, mixing bulk generators and generators of $U(1)$ gauge
groups on the branes, lead to Freed-Witten like requirements,
constraining the allowed brane configurations.

Comparison with ${\cal N}=4$ gauged supergravity theories allowed
us to establish a dictionary between fluxes and gaugings and to
relate Jacobi identities to quadratic constraints on gaugings,
necessary for consistency of supergravity theories. In particular,
some of these constraints appear to be stronger than those
previously discussed in the literature. There appear to be also
extra requirements that are not captured by the $\mathcal{N}=4$
algebra and which can be associated to D3-brane charges. It would be interesting to see if these extra
requirements could be obtained as Jacobi identities of an
extended algebra, presumably, associated to ${\cal N}=1$ gauged
supergravity theory. This fact deserves further investigation.

There are several other directions which this work leaves open.
For instance, we have not addressed the exploration of the vacuum
structure associated to the algebra (\ref{algfinal}). Some first
steps along this direction has been taken in
\cite{anamaria,guarino}, using techniques of computational
algebraic geometry. We find it would be interesting to know how
the results of those investigations are modified by the stronger
constraints we found here, or by including in the algebra the
gauge generators corresponding to the open string/twisted sector.
In particular, it is not transparent to us if solutions,
beyond the family we have described in section \ref{secfth} (
related to standard type IIB compactifications with 3-form fluxes
by four T-dualities and a S-duality rotation), do exist.

It has also been also shown in \cite{tdual} that generalized geometry
can provide a local geometrical definition of the NSNS flux
parameters appearing in the four dimensional effective gauged
supergravity. Here we have presented a procedure to extend these
algebras in order to account also for the RR degrees of freedom.
Definitively, it would be interesting to understand the new flux
parameters appearing in these extended algebras from the point of
view of exceptional generalized geometry \cite{exgener}.

We believe that the combination of general flux backgrounds, like
the ones considered here, with suitable chiral configurations of
F-theory 7-branes, is a good road towards the construction of
phenomenological models, providing powerful tools for addressing
not only the right spectrum of particles, but also moduli
stabilization or supersymmetry breaking. Hopefully, the algebras
we have presented in this work will serve to construct and analyze
phenomenological F-theory compactifications to four dimensions in
the near future.

\vspace*{1cm}
\newpage

{\bf \large Acknowledgments}

We thank E. Andr\'es, L. Carlevaro, A. Font, A. Guarino and F.
Marchesano for useful discussions and comments and, in particular,
A. Uranga who participated in the preliminary steps of this work.
The work of P.G.C. is supported by the European Union through an
Individual Marie-Curie IEF. Additional support comes from the
contracts ANR-05-BLAN-0079-02, MRTN-CT-2004-005104,
MRTN-CT-2004-503369, MEXT-CT-2003-509661 and CNRS PICS \#~2530,
3059, 3747. G.A. work is partially supported by   PIP5231
(CONICET)and  grant 06/C225 (UNC). A.R. thanks ICTP for partial support.

\newpage

\appendix

\section{Background fluxes}
\label{appA}

Here we collect tables containing the RR and NSNS flux parameters
surviving the $\mathbb{Z}_2\times \mathbb{Z}_2$ and orientifold
projections, and their tensorial structure in different T-dual
formulations, in the notation of \cite{acfi}.

\begin{table}[!ht] \footnotesize
\renewcommand{\arraystretch}{1.25}
\begin{center}
\begin{tabular}{cccc|cccc}
IIB/O3 & IIA/O6 & IIB/O9 & flux & IIB/O3 & IIA/O6 & IIB/O9 & flux \\
\hline
$F_{123}=-\tilde{F}^{456}$ & $F_0$ & $\! \! \! -\bF_{456}$ & $-m$ & $H_{123}=-\tilde{H}^{456}$ & $R^{123}$ & $\bR^{123}$ & $\ \ \bh_0$\\
$F_{423}=\tilde{F}^{156}$ & $F_{14}$ & $\bF_{156}$ & $-q_1$ & $H_{423}=\tilde{H}^{156}$ & $-Q_4^{23}$ & $\bR^{423}$ & $-\ba_1$\\
$F_{153}=\tilde{F}^{426}$ & $F_{25}$ & $\bF_{426}$ & $-q_2$ & $H_{153}=\tilde{H}^{426}$ & $-Q_5^{31}$ & $\bR^{153}$ & $-\ba_2$\\
$F_{126}=\tilde{F}^{453}$ & $F_{36}$ & $\bF_{453}$ & $-q_3$ & $H_{126}=\tilde{H}^{453}$ & $-Q_6^{12}$ & $\bR^{126}$ & $-\ba_3$\\
$F_{156}=-\tilde{F}^{423}$ & $F_{2536}$ & $\! \! \! -\bF_{423}$ & $\ \ e_1$ & $H_{156}=-\tilde{H}^{423}$ & $-\om^1_{56}$ & $\bR^{156}$ & $-a_1$\\
$F_{426}=-\tilde{F}^{153}$ & $F_{1436}$ & $\! \! \! -\bF_{153}$ & $\ \ e_2$ & $H_{426}=-\tilde{H}^{153}$ & $-\om^2_{64}$ & $\bR^{426}$ & $-a_2$\\
$F_{453}=-\tilde{F}^{126}$ & $F_{1425}$ & $\! \! \! -\bF_{126}$ & $\ \  e_3$ & $H_{453}=-\tilde{H}^{126}$ & $-\om^3_{45}$ & $\bR^{453}$ & $-a_3$ \\
$F_{456}=\tilde{F}^{123}$ & $F_{142536}$ & $\bF_{123}$ & $-e_0$ &
$H_{456}=\tilde{H}^{123}$ & $\ov{H}_{456}$ & $\bR^{456}$ & $\ \
h_0$
\end{tabular}
\end{center}
\caption{\small Type IIB/O3 3-form RR (left) and NSNS (right)
fluxes and their T-duals.} \label{rriib}
\end{table}

\begin{table}[!ht] \footnotesize
\renewcommand{\arraystretch}{1.25}
\begin{center}
\begin{tabular}{cccc}
IIB/O3 & IIA/O6 & IIB/O9 & flux \\
\hline $\bmat{ccc} \! \! \! Q^{23}_4 & Q^{31}_5 & Q^{12}_6 \! \!
\!\emat$ & $ \! \! -\bmat{ccc} \! \! \!  \ov{H}_{423} &
\ov{H}_{153} & \ov{H}_{126}  \! \! \! \emat$ & $\bmat{ccc} \! \!
\! \omega^4_{23} & \omega^5_{31} & \omega^6_{12}  \! \! \! \emat$
&
$ \! \! -\bmat{ccc} \! \! \!  h_1 & h_2 & h_3  \! \! \!\emat$  \\[0.35cm]
$\bmat{ccc} \! \! \!
-Q^{23}_1 & \, Q^{34}_5 & \, Q^{42}_6 \\
\, Q^{53}_4 & \! \! \! -Q^{31}_2 & \, Q^{15}_6 \\
\, Q^{26}_4 & \, Q^{61}_5 & \! \! \! -Q^{12}_3 \emat $  &
$\bmat{ccc} \! \! \!
-\om^1_{23} & \, \om^4_{53} & \, \om^4_{26} \\
\, \om^5_{34} & \! \! \! -\om^2_{31} & \, \om^5_{61} \\
\, \om^6_{42} & \, \om^6_{15} & \! \! \! -\om^3_{12} \emat$ &
$\bmat{ccc} \! \! \!
-\om^1_{23} & \, \om^5_{34} & \, \om^6_{42} \\
\, \om^4_{53} & \! \!\! -\om^2_{31} & \, \om^6_{15} \\
\, \om^4_{26} & \, \om^5_{61} & \!\! \! -\om^3_{12} \emat$ &
$\bmat{ccc} b_{11} & b_{12} & b_{13} \\
b_{21} & b_{22} & b_{23} \\
b_{31} & b_{32} & b_{33} \emat$ \\
& & & \\[-0.4cm]
$\bmat{ccc} \! \! \! Q^{56}_1 & Q^{64}_2 & Q^{45}_3  \! \! \!
\emat$ & $ \! \! -\bmat{ccc} \! \! \!  R^{156} & R^{426} & R^{453}
\! \! \! \emat$ & $\bmat{ccc} \! \! \! \om^1_{56} & \om^2_{64} &
\om^3_{45}  \! \! \! \emat$ &
$ \! \! -\bmat{ccc} \! \! \!  \bh_1 & \bh_2 & \bh_3  \! \! \! \emat$  \\[0.35cm]
$\bmat{ccc} \! \! \!
-Q^{56}_4 & \, Q^{61}_2 & \, Q^{15}_3 \\
\, Q^{26}_1 & \! \! \! -Q^{64}_5 & \, Q^{42}_3 \\
\, Q^{53}_1 & \, Q^{34}_2 & \! \! \! -Q^{45}_6 \emat$ &
$\bmat{ccc} \! \! \!
-Q^{56}_4 & \, Q^{26}_1 & \, Q^{53}_1 \\
\, Q^{61}_2 & \! \! \! -Q^{64}_5 & \, Q^{34}_2 \\
\, Q^{15}_3 & \, Q^{42}_3 & \! \! \! -Q^{45}_6 \emat$ &
$\bmat{ccc} \! \! \!
-\om^4_{56} & \, \om^2_{61} & \, \om^3_{15} \\
\, \om^1_{26} & \! \! \! -\om^5_{64} & \, \om^3_{42} \\
\, \om^1_{53} & \, \om^2_{34} & \! \! \! -\om^6_{45} \emat$ &
$\bmat{ccc}
\bb_{11} & \bb_{12} & \bb_{13} \\
\bb_{21} & \bb_{22} & \bb_{23} \\
\bb_{31} & \bb_{32} & \bb_{33} \emat$
\end{tabular}
\end{center}
\caption{\small Type IIB/O3 NSNS Q-fluxes and their T-duals.}
\label{nongeoiib}
\end{table}

\begin{table}[!ht] \footnotesize
\renewcommand{\arraystretch}{1.25}
\begin{center}
\begin{tabular}{cccc}
IIB/O3 & IIA/O6 & IIB/O9 & flux \\
\hline $\bmat{ccc} \! \! \! P^{23}_4 & P^{31}_5 & P^{12}_6 \! \!
\!\emat$ & $ \! \! -\bmat{ccc} \! \! \!  P_{423} & P_{153} &
P_{126}  \! \! \! \emat$ & $\bmat{ccc} \! \! \! P^4_{23} &
P^5_{31} & P^6_{12}  \! \! \! \emat$ &
$ \! \! -\bmat{ccc} \! \! \!  f_1 & f_2 & f_3  \! \! \!\emat$  \\[0.35cm]
$\bmat{ccc} \! \! \!
-P^{23}_1 & \, P^{34}_5 & \, P^{42}_6 \\
\, P^{53}_4 & \! \! \! -P^{31}_2 & \, P^{15}_6 \\
\, P^{26}_4 & \, P^{61}_5 & \! \! \! -P^{12}_3 \emat $  &
$\bmat{ccc} \! \! \!
-P^1_{23} & \, P^4_{53} & \, P^4_{26} \\
\, P^5_{34} & \! \! \! -P^2_{31} & \, P^5_{61} \\
\, P^6_{42} & \, P^6_{15} & \! \! \! -P^3_{12} \emat$ &
$\bmat{ccc} \! \! \!
-P^1_{23} & \, P^5_{34} & \, P^6_{42} \\
\, P^4_{53} & \! \!\! -P^2_{31} & \, P^6_{15} \\
\, P^4_{26} & \, P^5_{61} & \!\! \! -P^3_{12} \emat$ &
$\bmat{ccc} g_{11} & g_{12} & g_{13} \\
g_{21} & g_{22} & g_{23} \\
g_{31} & g_{32} & g_{33} \emat$\\
& & & \\[-0.4cm]
$\bmat{ccc} \! \! \! P^{56}_1 & P^{64}_2 & P^{45}_3  \! \! \!
\emat$ & $ \! \! -\bmat{ccc} \! \! \!  P^{156} & P^{426} & P^{453}
\! \! \! \emat$ & $\bmat{ccc} \! \! \! P^1_{56} & P^2_{64} &
P^3_{45}  \! \! \! \emat$ &
$ \! \! -\bmat{ccc} \! \! \!  \bar f_1 & \bar f_2 & \bar f_3  \! \! \! \emat$  \\[0.35cm]
$\bmat{ccc} \! \! \!
-P^{56}_4 & \, P^{61}_2 & \, P^{15}_3 \\
\, P^{26}_1 & \! \! \! -P^{64}_5 & \, P^{42}_3 \\
\, P^{53}_1 & \, P^{34}_2 & \! \! \! -P^{45}_6 \emat$ &
$\bmat{ccc} \! \! \!
-P^{56}_4 & \, P^{26}_1 & \, P^{53}_1 \\
\, P^{61}_2 & \! \! \! -P^{64}_5 & \, P^{34}_2 \\
\, P^{15}_3 & \, P^{42}_3 & \! \! \! -P^{45}_6 \emat$ &
$\bmat{ccc} \! \! \!
-P^4_{56} & \, P^2_{61} & \, P^3_{15} \\
\, P^1_{26} & \! \! \! -P^5_{64} & \, P^3_{42} \\
\, P^1_{53} & \, P^2_{34} & \! \! \! -P^6_{45} \emat$ &
$\bmat{ccc}
\bg_{11} & \bg_{12} & \bg_{13} \\
\bg_{21} & \bg_{22} & \bg_{23} \\
\bg_{31} & \bg_{32} & \bg_{33} \emat$
\end{tabular}
\end{center}
\caption{\small Type IIB/O3 RR P-fluxes and their T-duals.}
\label{pfluxes}
\end{table}

\noindent We also state the superpotential (\ref{super}), in terms
of the flux parameters defined in table \ref{mapa},
\beqa W_{Flux} & = & e_0-i\sum_{i=1}^3
h_iT_i+\frac{1}{2}\sum_{l\aneq m\aneq
n} h'_lT_mT_n+i e'_0 T_1T_2T_3 \label{todo} \\
& + & \bigg(ih_0-\sum_{i=1}^3 f_iT_i-\frac{i}{2}\sum_{l\aneq
m\aneq
n} f'_lT_mT_n - h'_0 T_1T_2T_3 \bigg)S \nonumber \\
& + & \sum_{i=1}^3 \left[\bigg(-a_i+i\sum_{j=1}^3g_{ij}T_j
-\frac{1}{2}\sum_{l\aneq m\aneq n} g'_{il}T_mT_n + ia'_i T_1T_2T_3 \bigg)S \right. \nonumber \\
& + & \left.  ie_i-\sum_{j=1}^3 b_{ij}T_j-\frac{i}{2}\sum_{l\aneq
m\aneq n}b'_{il}T_mT_n
-e'_iT_1T_2T_3 \right] U_i \nonumber\\
& + &  \frac{1}{2}\sum_{r\aneq s\aneq
t}\left[\bigg(i\bar{a}_r+\sum_{j=1}^3\bar{g}_{rj}T_j+\frac{i}{2}
\sum_{l\aneq m\aneq n} \bar{g}'_{rl}T_mT_n - \bar{a}'_rT_1T_2T_3 \bigg)S \right. \nonumber\\
& - &  \left. q_r+i\sum_{j=1}^3\bar{b}_{rj}T_j
-\frac{1}{2}\sum_{l\aneq m\aneq n}\bar{b}'_{rl}T_mT_n + iq'_r
T_1T_2T_3\right]U_sU_t \nonumber \\
& + &
\left[-\bigg(\bar{h}_0+i\sum_{j=1}^3\bar{f}_{j}T_j-\frac{1}{2}\sum_{l\aneq
m\aneq n}\bar{f}'_lT_mT_n + i\bar{h}'_0T_1T_2T_3\bigg)S \right. \nonumber\\
& + &  \left.
im+\sum_{j=1}^3\bar{h}_{j}T_j+\frac{i}{2}\sum_{l\aneq m\aneq
n}\bar{h}'_lT_mT_n- m'T_1T_2T_3\right]U_1U_2U_3 \nonumber \eeqa

\begin{table}[!ht]
\begin{center}
\begin{tabular}{c|c||c|c}
Flux parameter & Weight & Flux parameter & Weight\\
\hline \hline $\bar{h}'_0$ & $(-,-,-,-,-,-,-)$& $e_0$ & $(+,+,+,+,+,+,+)$\\
$h_0$ & $(-,+,+,+,+,+,+)$& $m'$ & $(+,-,-,-,-,-,-)$\\
$-h_i$ & $(+,{\stackrel{\mathsmaller{i}}{\overbrace{-,+,+}}},+,+,+)$ & $-\bar{f}'_i$ &
$(-,{\stackrel{\mathsmaller{i}}{\overbrace{+,-,-}}},-,-,-)$\\
$e_j$ & $(+,+,+,+,{\stackrel{\mathsmaller{j}}{\overbrace{-,+,+}}})$ & $\bar{a}'_j$ &
$(-,-,-,-,{\stackrel{\mathsmaller{j}}{\overbrace{+,-,-}}})$\\
$\bar{h}'_i$ &
$(+,{\stackrel{\mathsmaller{i}}{\overbrace{+,-,-}}},-,-,-)$&$f_i$ &
$(-,{\stackrel{\mathsmaller{i}}{\overbrace{-,+,+}}},+,+,+)$\\
$q'_j$ &
$(+,-,-,-,{\stackrel{\mathsmaller{j}}{\overbrace{+,-,-}}})$&$a_j$ &
$(-,+,+,+,{\stackrel{\mathsmaller{j}}{\overbrace{-,+,+}}})$\\
$\bar{g}'_{ji}$ &
$(-,{\stackrel{\mathsmaller{i}}{\overbrace{+,-,-}}},{\stackrel{\mathsmaller{j}}{\overbrace{+,-,-}}})$&
$b_{ji}$ & $(+,{\stackrel{\mathsmaller{i}}{\overbrace{-,+,+}}},{\stackrel{\mathsmaller{j}}{\overbrace{-,+,+}}})$\\
$a'_j$ &
$(-,-,-,-,{\stackrel{\mathsmaller{j}}{\overbrace{-,+,+}}})$&$q_j$ &
$(+,+,+,+,{\stackrel{\mathsmaller{j}}{\overbrace{+,-,-}}})$\\
$-g_{ji}$ &
$(-,{\stackrel{\mathsmaller{i}}{\overbrace{-,+,+}}},{\stackrel{\mathsmaller{j}}{\overbrace{-,+,+}}})$&
$-\bar{b}'_{ji}$ &
$(+,{\stackrel{\mathsmaller{i}}{\overbrace{+,-,-}}},{\stackrel{\mathsmaller{j}}{\overbrace{+,-,-}}})$\\
$-\bar{a}_j$&$(-,+,+,+,{\stackrel{\mathsmaller{j}}{\overbrace{+,-,-}}})$&
$-e'_j$&$(+,-,-,-,{\stackrel{\mathsmaller{j}}{\overbrace{-,+,+}}})$\\
$-\bar{b}_{ji}$&
$(+,{\stackrel{\mathsmaller{i}}{\overbrace{-,+,+}}},{\stackrel{\mathsmaller{j}}{\overbrace{+,-,-}}})$&
$-g'_{ji}$&$(-,{\stackrel{\mathsmaller{i}}{\overbrace{+,-,-}}},{\stackrel{\mathsmaller{j}}{\overbrace{-,+,+}}})$\\
$-m$&$(+,+,+,+,-,-,-)$&$-h'_0$&$(-,-,-,-,+,+,+)$\\
$b'_{ji}$&
$(+,{\stackrel{\mathsmaller{i}}{\overbrace{+,-,-}}},{\stackrel{\mathsmaller{j}}{\overbrace{-,+,+}}})$&
$\bar{g}_{ji}$&
$(-,{\stackrel{\mathsmaller{i}}{\overbrace{-,+,+}}},{\stackrel{\mathsmaller{j}}{\overbrace{+,-,-}}})$\\
$f'_i$&
$(-,{\stackrel{\mathsmaller{i}}{\overbrace{+,-,-}}},+,+,+)$&
$\bar{h}_i$&$(+,{\stackrel{\mathsmaller{i}}{\overbrace{-,+,+}}},-,-,-)$\\
$-e'_0$&$(+,-,-,-,+,+,+)$&$-\bar{h}_0$&$(-,+,+,+,-,-,-)$\\
$-\bar{f}_i$&$(-,{\stackrel{\mathsmaller{i}}{\overbrace{-,+,+}}},-,-,-)$&
$-h'_i$&$(+,{\stackrel{\mathsmaller{i}}{\overbrace{+,-,-}}},+,+,+)$
\end{tabular}
\end{center}
\caption{Spinorial embedding of the background fluxes. The weights in each column correspond to
one of the two Weyl spinors on which the set of fluxes ${\mathbb G}$ can be decomposed.\vspace{2cm}} \label{mapa}
\end{table}

\newpage

{\small

\end{document}